\documentclass[twocolumn]{aastex6}
\usepackage{times}
\usepackage{amsmath}
\usepackage{textcomp}
\usepackage{amssymb}
\usepackage{natbib}
\usepackage{float}
\usepackage{color}
\usepackage{graphicx}
\usepackage{epstopdf}
\newcommand{\Msun}{\ensuremath{M_{\odot}}}
\newcommand{\lum}{erg\,s$^{-1}$}
\newcommand{\fermi}{{\it Fermi}}
\newcommand{\nustar}{{\it NuSTAR}}

\newcommand{\swift}{{\it Swift}}

\newcommand{\gm}{$\gamma$}
\slugcomment{submitted to ApJ}

\shorttitle{BAT Blazars}
\shortauthors{Paliya et al.}

\begin{document}
\title{BAT AGN Spectroscopic Survey: XVI. General Physical Characteristics of BAT Blazars}

\author{Vaidehi S. Paliya$^{1}$, M. Koss$^2$, B. Trakhtenbrot$^3$, C. Ricci$^{4,5}$, K. Oh$^{6,7}$, M. Ajello$^{8}$, D. Stern $^9$, M. C. Powell$^{10}$, C. M. Urry$^{10}$, F. Harrison$^{11}$, I. Lamperti$^{12}$, R. Mushotzky$^{13}$, L. Marcotulli$^{8}$,  J. Mej\'ia-Restrepo$^{14}$, and D. Hartmann$^{8}$} 
\affil{$^1$Deutsches Elektronen Synchrotron DESY, Platanenallee 6, 15738 Zeuthen, Germany}
\affil{$^2$Eureka Scientific Inc, Oakland, CA, 94602, USA}
\affil{$^3$School of Physics and Astronomy, Tel Aviv University, Tel Aviv 69978, Israel} 
\affil{$^4$N\'ucleo de Astronom\'ia de la Facultad de Ingenier\'ia, Universidad Diego Portales, Av. Ej\'ercito Libertador 441, Santiago, Chile}
\affil{$^5$Kavli Institute for Astronomy and Astrophysics, Peking University, Beijing 100871, China}
\affil{$^6$Department of Astronomy, Kyoto University, Oiwake-cho, Sakyo-ku, Kyoto 606-8502, Japan}
\affil{$^7$JSPS Fellow}
\affil{$^{8}$Department of Physics and Astronomy, Clemson University, Kinard Lab of Physics, Clemson, SC 29634-0978, USA}
\affil{$^9$Jet Propulsion Laboratory, California Institute of Technology, 4800 Oak Grove Drive, MS 169-224, Pasadena, CA 91109, USA}
\affil{$^{10}$Yale Center for Astronomy and Astrophysics, and Physics Department, Yale University, P.O. Box 2018120, New Haven, CT 06520-8120, USA}
\affil{$^{11}$Cahill Center for Astronomy and Astrophysics, California Institute of Technology, Pasadena, CA 91125, USA}
\affil{$^{12}$Department of Physics \& Astronomy, University College London, Gower Street, London, WC1E 6BT, UK}
\affil{$^{13}$Department of Astronomy and Joint Space-Science Institute, University of Maryland, College Park, MD 20742, USA}
\affil{$^{14}$European Southern Observatory, Alonso de Cordova 3107, Casilla 19001, Victacura, Santiago, Chile}

\email{vaidehi.s.paliya@gmail.com}

\begin{abstract}
The recently released 105-month \swift-Burst Alert Telescope (BAT) all-sky hard X-ray survey catalog presents an opportunity to study astrophysical objects detected in the deepest look at the entire hard X-ray (14$-$195 keV) sky. Here we report the results of a multifrequency study of 146 blazars from this catalog, quadrupling the number compared to past studies, by utilizing recent data from the \fermi-Large Area Telescope (LAT), \swift-BAT, and archival measurements. In our \gm-ray analysis of $\sim$10 years of the LAT data, 101 are found as \gm-ray emitters, whereas, 45 remains LAT undetected. We model the broadband spectral energy distributions with a synchrotron-inverse Compton radiative model. On average, BAT detected sources host massive black holes ($M_{\rm bh}\sim10^9$ \Msun) and luminous accretion disks ($L_{\rm d}\sim10^{46}$ \lum). At high-redshifts ($z>2$), BAT blazars host more powerful jets with luminous accretion disks compared to those detected only with the \fermi-LAT. We find good agreement in the black hole masses derived from the single-epoch optical spectroscopic measurements and standard accretion disk modeling approaches. Other physical properties of BAT blazars are similar to those known for \fermi-LAT detected objects.

\end{abstract}


\section{Introduction}{\label{sec:Intro}}
Active galactic nuclei (AGN) are among the most energetic objects in the Universe and are crucial players in the evolution of galaxies \citep[see, e.g.][for a review]{2013ARA&A..51..511K}. A subset of these interesting objects host relativistic jets and when the jet is closely aligned with the line of sight to the observer, the source is termed as a blazar \citep[][]{1978PhyS...17..265B}. Due to their peculiar orientation, the emitted radiation from blazar jets is relativistically amplified. This phenomenon makes blazars visible even at very high-redshifts \citep[$z>2$, e.g.][]{2004ApJ...610L...9R,2017ApJ...837L...5A}. Blazars are classified as flat spectrum radio quasars (FSRQs) and/or BL Lac objects based on the rest-frame equivalent width (EW) of the optical emission lines \citep[][]{1991ApJ...374..431S} with FSRQs exhibiting broad emission lines (EW$>$5\AA), while BL Lacs show weak or no emission lines in their optical spectra. This, in turn, suggests a radiatively efficient accretion process that illuminates the broad line region (BLR) surrounding the central engine of FSRQs \citep[cf.][]{2011MNRAS.414.2674G}. The spectral energy distribution (SED) of a blazar is characterized by a double hump structure. The low energy peak is understood to be due to synchrotron process and is typically observed between the infrared and the X-ray band, whereas, the high energy peak is typically explained by the inverse Compton scattering of the low energy photons by relativistic electrons in the jet. Alternatively, hadronic models have also been invoked to explain the high energy SEDs of blazars \citep[e.g.][]{2013ApJ...768...54B}. Based on the location of the synchrotron peak frequency ($\nu^{\rm peak}_{\rm syn}$), blazars are also classified as low- ($\nu^{\rm peak}_{\rm syn}<$10$^{14}$ Hz), intermediate- (10$^{14}<\nu^{\rm peak}_{\rm syn}<$10$^{15}$ Hz) and high- ($\nu^{\rm peak}_{\rm syn}>$10$^{15}$ Hz) synchrotron peaked or LSP, ISP, and HSP, respectively \citep[][]{2010ApJ...716...30A}. Typically FSRQs are LSP/ISP blazars, whereas, BL Lacs are mostly HSP ones.

The sky-surveying capabilities of the currently operating hard X-ray and \gm-ray instruments, namely Burst Alert Telescope \citep[BAT;][]{2005SSRv..120..143B} onboard {\it Neil Gehrels Swift} observatory \citep[][]{2004ApJ...611.1005G} and Large Area Telescope \citep[LAT;][]{2009ApJ...697.1071A} onboard \fermi~Gamma-ray Space Telescope, have allowed to carry out the population studies of blazars and their luminosity dependent evolution \citep[e.g.][]{2009ApJ...699..603A,2012ApJ...751..108A,2014ApJ...780...73A}. In particular, \citet[][]{2009ApJ...699..603A} studied 38 blazars detected in the first 36 months of the all-sky survey by \swift-BAT and used this sample to derive the 15$-$55 keV luminosity function (LF) of blazars. Interestingly, it was noted that the evolution of luminous FSRQs peaks at higher redshifts ($\sim$4) compared to other classes of BAT detected AGNs. \citet[][]{2010MNRAS.405..387G} used the LF reported in \citet[][]{2009ApJ...699..603A} to determine the space density of billion-solar-mass black holes residing in jetted systems and introduced an exponential cut-off in the blazar luminosity function above $z=4.3$ to remain consistent with the number of detected blazars in that redshift bin and also with the space density of massive halos. However, computation of the luminosity function depends strongly on the number of sources detected in a particular redshift bin. For example, \citet[][]{2017ApJ...837L...5A} reported the first time \gm-ray detection of five $z>3.1$ radio-loud quasars thus confirming their blazar nature. This allowed them to update the space density of 10$^9$ \Msun~black holes residing in radio-loud quasars at $z\approx4$ and to conclude that radio-loudness may be a crucial ingredient for the rapid black hole growth in the early Universe \citep[see also][]{2011MNRAS.416..216V}. Moreover, by studying high-redshift ($z>2$) \swift-BAT blazars, \citet[][]{2010MNRAS.405..387G} found them to host more powerful jets and more luminous accretion disks compared to \fermi-LAT detected $z>2$ sources.

A new catalog of \swift-BAT detected sources covering the first 105-month of the mission has recently been released \citep[][]{2018ApJS..235....4O} along with an extensive multi-wavelength observations of BAT detected sources by the BAT AGN Spectroscopic Survey (BASS) collaboration \citep[][]{2017ApJ...850...74K,2017MNRAS.467..540L,2017ApJS..233...17R}. Motivated by more than quadrupling of the sources compared to past studies and also the larger redshift range it covers\footnote{The farthest known blazar in 105-month catalog, SWIFT J1430.6+4211, has $z=4.71$, significantly larger than the farthest known sources in previous BAT catalogs \citep[$z=3.67$, e.g.,][]{2009ApJ...699..603A}.}, we have carried out a multiwavelength study of all the blazars present in the 105-month \swift-BAT catalog. Our primary objective is to explore the physical properties of \swift-BAT blazars by applying a simple leptonic radiative model and also to compare them with that known for the \fermi-LAT detected ones which are yet to be detected by BAT. 

 In Section~\ref{sec2}, we describe the sample adopted in this work and data analysis procedures are explained in Section~\ref{sec3}. The adopted leptonic model is elaborated in Section~\ref{sec4} and we discuss the derived SED parameters in Section~\ref{sec5}. In Section~\ref{sec6}, we compare the accretion-jet connection observed in BAT blazars with \fermi-LAT detected sources, and Section~\ref{sec7} is devoted to the luminosity dependent evolution of blazars. We summarize our findings in Section~\ref{sec8}. 
 Throughout this work, we use a flat cosmology with $H_0=67.8$ km s$^{-1}$ Mpc$^{-1}$ and $\Omega_{\rm M}=0.308$ \citep[][]{2016A&A...594A..13P}.
 
\section{The Sample}\label{sec2}
Our blazar sample is based on the 1632 sources included in the 105-month \swift-BAT catalog. There are 158 sources classified as `beamed AGN', mainly based on their presence in the BZCAT and CRATES catalogs \citep[][]{2015Ap&SS.357...75M,2007ApJS..171...61H}. We exclude those 32 objects which are reported as `beamed AGN' (based on the BZCAT catalog) but, however, do not exhibit broadband properties typically observed from the blazar class of AGN. The full list of these sources is provided in Table~\ref{tab:excluded}. Additionally, we explore the broadband properties of the whole BAT sample to identify potential blazar candidates which might have been misclassified as Seyferts. This was done by examining their multifrequency SEDs generated using the ASDC SED builder tool\footnote{https://tools.asdc.asi.it/} and physical properties (e.g., radio detection) reported in literature \citep[cf.][]{1993AJ....105.1666G,1996ApJS..103..427G}. This led to the inclusion of 20 sources (Table~\ref{tab:added}). Admittedly, there could be a very few unidentified BAT objects that might be blazars though it may not be possible to pinpoint them due to lack of a confirmed optical/radio counterparts and also because of paucity of the multiwavelength information. The BAT survey completeness and redshift evolution of beamed AGN will be fully examined in a future study (Marcotulli et al. in prep.).

This paper makes use of redshift and black hole mass estimates from BASS, a large effort to collect optical spectra for all the \swift-BAT AGN.  The data release 1 optical spectra were obtained from a large variety of telescopes \citep[][]{2017ApJ...850...74K}, and can all be viewed at the BASS website\footnote{www.bass-survey.com}. We also include 15 new Data Release 2 redshifts measurements, from Palomar, Southern Astrophysical Research (SOAR) telescope, and the Very Large Telescope (VLT) that will shortly be published (Oh et al., in prep) as well as comparison black hole mass measuremenst from this catalog. Altogether, our final sample consists of 146 BAT blazars. 

\section{Data Reduction and Compilation}\label{sec3}
\subsection{\fermi-LAT}
The primary objective of this study is to explore the average physical characteristics of BAT blazars rather than any of their specific activity states. With this in mind, we perform an analysis of \fermi-LAT data covering $\sim$10 years (2008 August 5 to 2018 March 3) of \fermi~operation. The all-sky surveying capability of \fermi-LAT ensures that the generated spectra represent an average activity state of the blazar. A standard data reduction procedure is adopted to carry out the analysis\footnote{http://fermi.gsfc.nasa.gov/ssc/data/analysis/documentation/}. The data reduction is performed using Science Tools (v11r5p3) and publicly available analysis package {\tt fermiPy} \citep[][]{2017arXiv170709551W}. We use the latest P8R3 dataset \citep[][]{2013arXiv1303.3514A,2018arXiv181011394B} in the energy range of 0.1$-$300 GeV and consider only SOURCE class events ({\tt evclass=128}). We define a ROI of 15$^{\circ}$ radius centered at the target blazar and apply standard cuts (zenith angle $z_{\rm max}<90^{\circ}$ and {\tt `DATA\_QUAL$>$0 and LAT\_CONFIG==1'}) to select good time intervals. To derive the optimized spectral parameters from the binned-likelihood fitting, we consider all \gm-ray sources present in third \fermi-LAT detected source catalog \citep[3FGL;][]{2015ApJS..218...23A} and lying within the ROI and also the isotropic and Galactic diffuse emission models \citep[][]{2016ApJS..223...26A}. Note that the presence of any extended \gm-ray source in the ROI is properly taken into account by adopting the publicly available extended source templates\footnote{https://github.com/fermi-lat/extendedArchive}. The parameters of all of the considered \gm-ray sources are allowed to vary during the likelihood fitting. 

The significance of source detection is quantified using maximum likelihood test statistic TS =  2$\Delta \log \mathcal{L}$, where $\mathcal{L}$ denotes the likelihood ratio, between models with and without a point source at the position of interest. Since the time period covered is significantly larger than that used to produce the 3FGL catalog, we adopt {\tt findsource} tool available in {\tt fermiPy} to generate TS maps and search unmodeled \gm-ray sources that are present in the data \citep[with TS$\gtrsim$25, $\gtrsim$5$\sigma$ detection][]{1996ApJ...461..396M} but not in the model. Once found, they are characterized with a power law and included in the sky model. This procedure is repeated until the TS map stops showing any excess residual. Flux upper limits in the \gm-ray spectra are derived at 95\% confidence for energy bins with TS$<9$. Finally, we compute 5$\sigma$ LAT sensitivity limits by adopting a photon index of 2.4 for each \gm-ray undetected BAT FSRQ (median 0.1$-$300 GeV 5$\sigma$ flux upper limit $\sim$7$\times$10$^{-7}$ MeV cm$^{-2}$ s$^{-1}$). The generated \gm-ray spectra are corrected for extragalactic background light absorption following the prescriptions of \citet[][]{2011MNRAS.410.2556D}.

Overall, we found 101 \gm-ray emitting sources out of 146 BAT blazars and their derived \gm-ray spectral parameters are provided in Table~\ref{tab:fermi}.

\subsection{\swift-BAT, \nustar~and Archival Measurements}
We extract 105-month averaged 14$-$195 keV spectra for all sources using publicly available spectrum and response files and by fitting a simple power law model in XSPEC \citep[][]{1996ASPC..101...17A}, as done by \citet[][]{2018ApJS..235....4O}. 

The BAT spectra of 25 FSRQs exhibit a steep falling shape (photon index $>$2), in contrast to the soft X-ray ($<$10 keV) spectral shape which predicts a hard, rising (photon index $<$2) spectrum. This is likely due to low signal-to-noise spectrum in BAT where $\Gamma$ is not well constrained ($\Gamma_{Err}>0.4$). For such sources, we search for better quality \nustar~observations, when available, and find that in all cases \nustar~spectra matches well with the soft X-ray one. The data from the \nustar~focal plane modules A and B are reduced following the standard procedure. We first run the tool {\tt nupipeline} to clean and calibrate the event files. The source and background regions of 30$^{\prime\prime}$ and 70$^{\prime\prime}$, respectively, are then selected from the same chip, and the {\tt nuproducts} pipeline is adopted to extract the source and background spectra along with ancillary and response matrix files. The spectra are binned to have at least 20 counts per bin and fitting is performed in XSPEC.

Both \fermi-LAT and \swift-BAT are all-sky surveying instruments and hence considering their long-time averaged spectra is justified to achieve the objectives proposed in this work. On the other hand, soft X-ray and optical-UV instruments, e.g., X-ray telescope (XRT) and ultraviolet and optical telescope (UVOT) on board \swift, operate in pointed modes and observations are often triggered during elevated activity states of blazars. With this in mind, we do not separately analyze XRT and UVOT observations and rather rely on publicly available archival SED measurements from ASDC that can be considered as representatives of the average behavior of the target of interest. Note that the BASS collaboration is running a dedicated X-ray spectroscopic followup of BAT detected objects, irrespective of their activity states, and the results of a comprehensive an X-ray analysis can be found in \citet[][]{2017ApJS..233...17R}.

\section{Modeling the Broadband Emission}\label{sec4}
 \begin{figure*}[t!]
\hbox{\hspace{0cm}
\includegraphics[scale=0.34]{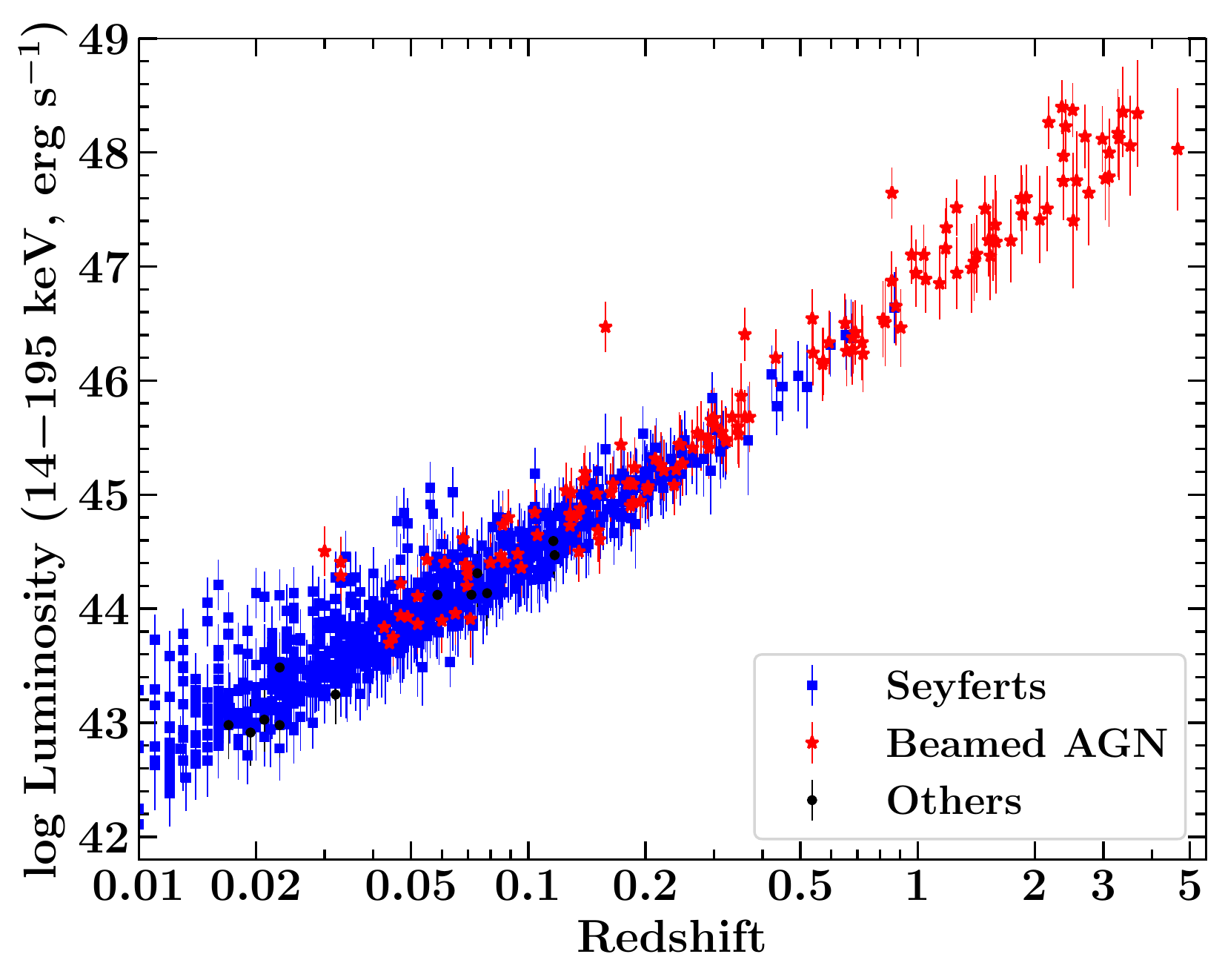}
\includegraphics[scale=0.34]{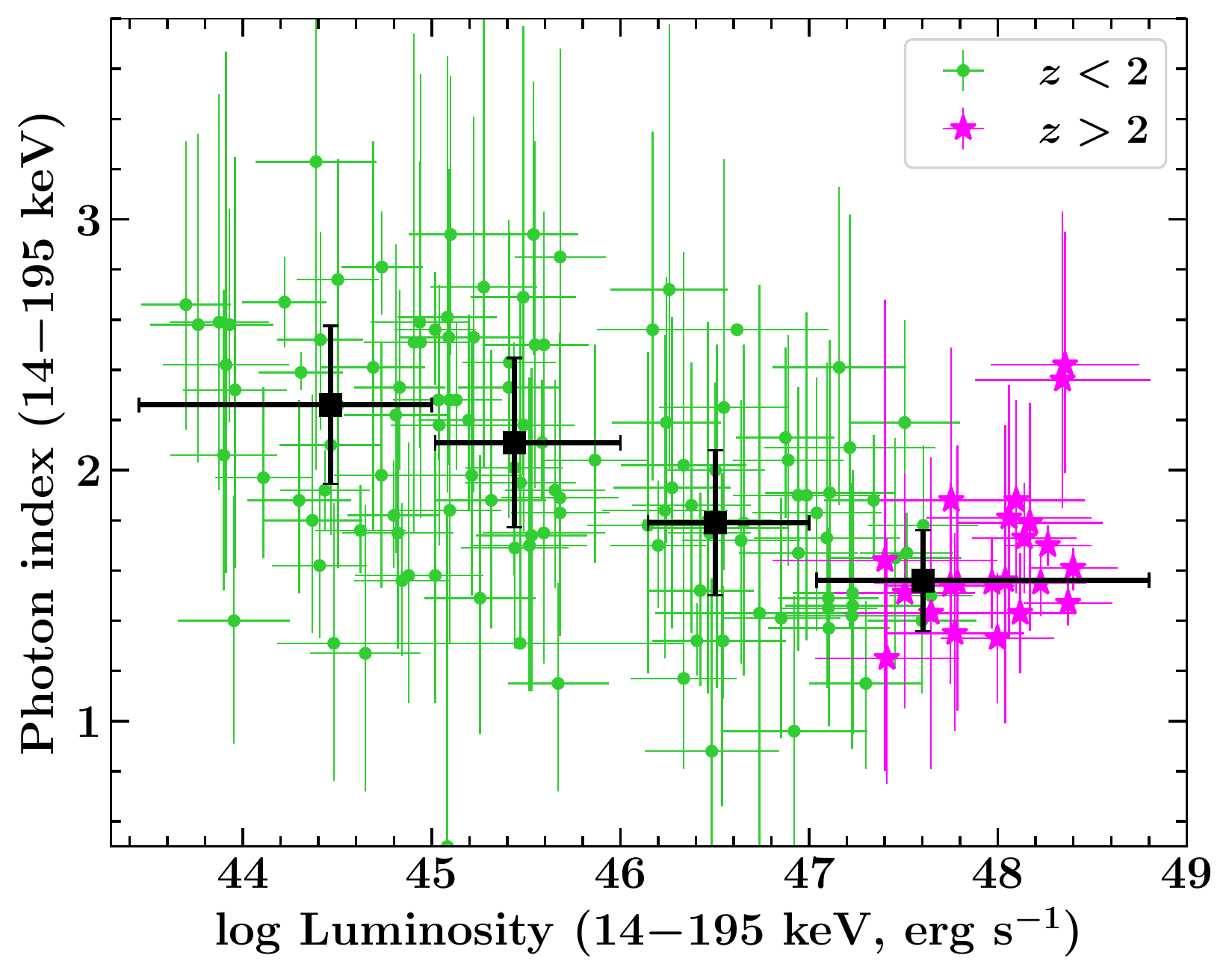}
\includegraphics[scale=0.34]{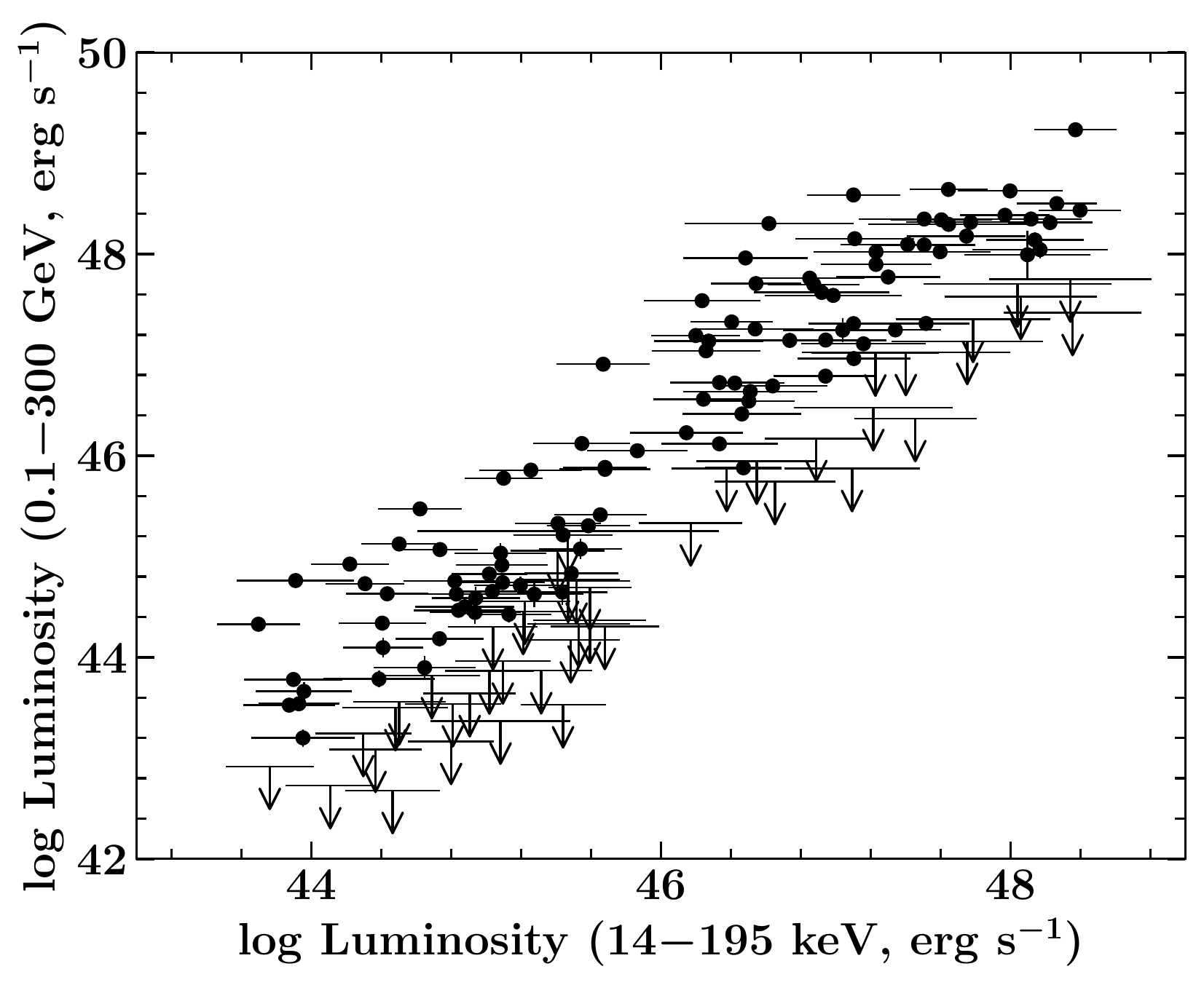}
}
\caption{Left: The redshift dependence of 14$-$195 keV luminosity for various types of BAT detected AGNs. Middle: A plot of the X-ray photon index versus luminosity. High-redshift blazars are displayed with pink stars. Black squares denote the median values of the plotted quantities in different luminosity bins and the uncertainties are median absolute deviation. Right: A plot of the hard X-ray versus \gm-ray luminosities. The downward arrows represent \gm-ray luminosity upper limits for \fermi-LAT undetected sources.} \label{fig:lumin}
\end{figure*}
\subsection{The Model}
 We adopt a simple one-zone leptonic emission model \citep[e.g.,][]{2009ApJ...692...32D,2009MNRAS.397..985G} to reproduce the multifrequency SEDs of BAT blazars. We consider a conical jet with a semi-opening angle of 0.1 radian. The emission region is assumed to be spherical and cover the whole cross-section of the jet. With this assumption, the dissipation distance ($R_{\rm d}$) constrains the size of the emitting region. The observer receives the Doppler-boosted jet emission at an angle of $\theta_{\rm v}$ from the jet. Furthermore, the radiating population of relativistic electrons has a broken power-law distribution of the following type:
 
 \begin{equation}
S(\gamma)  \, = \, S_0\, { (\gamma_{\rm b})^{-p} \over
(\gamma/\gamma_{\rm b})^{p} + (\gamma/\gamma_{\rm b})^{q}}
\end{equation}
where $S_0$ is the normalization constant (cm$^{-3}$), \gm~is the the Lorentz factor of electrons, and $p$, $q$ are the spectral indices below and above the break Lorentz factor $\gamma_{\rm b}$, respectively. Note that our modeling technique does not include a time-dependent evolution of the electron spectrum considering particle injection/cooling.

 The leptonic population emits synchrotron radiation in presence of uniform and randomly oriented magnetic field $B$. We also consider various photon fields, both inside and outside of the jet, to calculate the inverse Compton radiation from the emitting population \citep[e,g.,][]{1979rpa..book.....R}. This includes thermal photons originated from the accretion disk, BLR, and dusty torus \citep[the external Compton or EC process, e.g.,][]{1994ApJ...421..153S} and also non-thermal synchrotron photons produced by the same electrons \citep[synchrotron self Compton or SSC,][]{1985ApJ...298..114M}. Moreover, we assume the BLR and torus are spherical shells with radii $R_{\rm blr}=10^{17}\sqrt{L_{\rm d, 45}}$ cm and $R_{\rm torus}=2.5\times10^{18}\sqrt{L_{\rm d, 45}}$ cm, respectively, where $L_{\rm d,45}$ is the luminosity of the accretion disk in units of 10$^{45}$ \lum~\citep[][]{2009MNRAS.397..985G}. The adopted emission profiles of these components follow a blackbody distribution peaking at Hydrogen Ly-$\alpha$ frequency and the characteristic temperature of the torus ($T_{\rm torus}$), respectively. We assume a geometrically thin and optically thick accretion disk \citep[][]{1973A&A....24..337S} whose radiative profile is considered to follow a multi-colored blackbody \citep[see, e.g.,][]{2002apa..book.....F}

 \begin{equation}
 \label{eq:disk_flux}
 F_{\rm \nu,~disk} = \nu^3\frac{4\pi h \cos \theta_{\rm v} }{c^2 {D_l}^2}\int_{R_{\rm d,~in}}^{R_{\rm d,~out}}\frac{R\,{\rm d}R}{e^{h\nu/kT(R)}-1}
\end{equation}
where $h$ is the Planck constant, $D_l$ is the luminosity distance, $k$ is the Boltzmann constant, $c$ is the speed of light, and $R_{\rm d,~in}$ and $R_{\rm d,~out}$ are the inner and outer disk radii, assumed as 3$R_{\rm Sch}$ and 500$R_{\rm Sch}$, respectively. $R_{\rm Sch}$ is the Schwarzschild radius. In the above equation, the disk temperature has the following dependence on the radius

\begin{equation}
\label{eq:temp_profile}
T(R)\, =\, {  3 R_{\rm Sch}  L_{\rm d}  \over 16 \pi\eta_{\rm acc}\sigma_{\rm SB} R^3 }  
\left[ 1- \left( {3 R_{\rm Sch} \over  R}\right)^{1/2} \right]^{1/4},
\end{equation}
where $\sigma_{\rm SB}$ is the Stefan-Boltzmann constant and $\eta_{\rm acc}$ is the accretion efficiency, adopted here as 10\%. The thermal emission originated from the X-ray corona is assumed to follow a power law with exponential cut-off \citep[$L_{\rm cor}(\nu)  \propto \nu^{-1}  \exp (- h\nu / 150\, {\rm keV} )$, see][]{2009MNRAS.397..985G}.

\begin{figure*}[t!]
\hbox{
\includegraphics[scale=0.7]{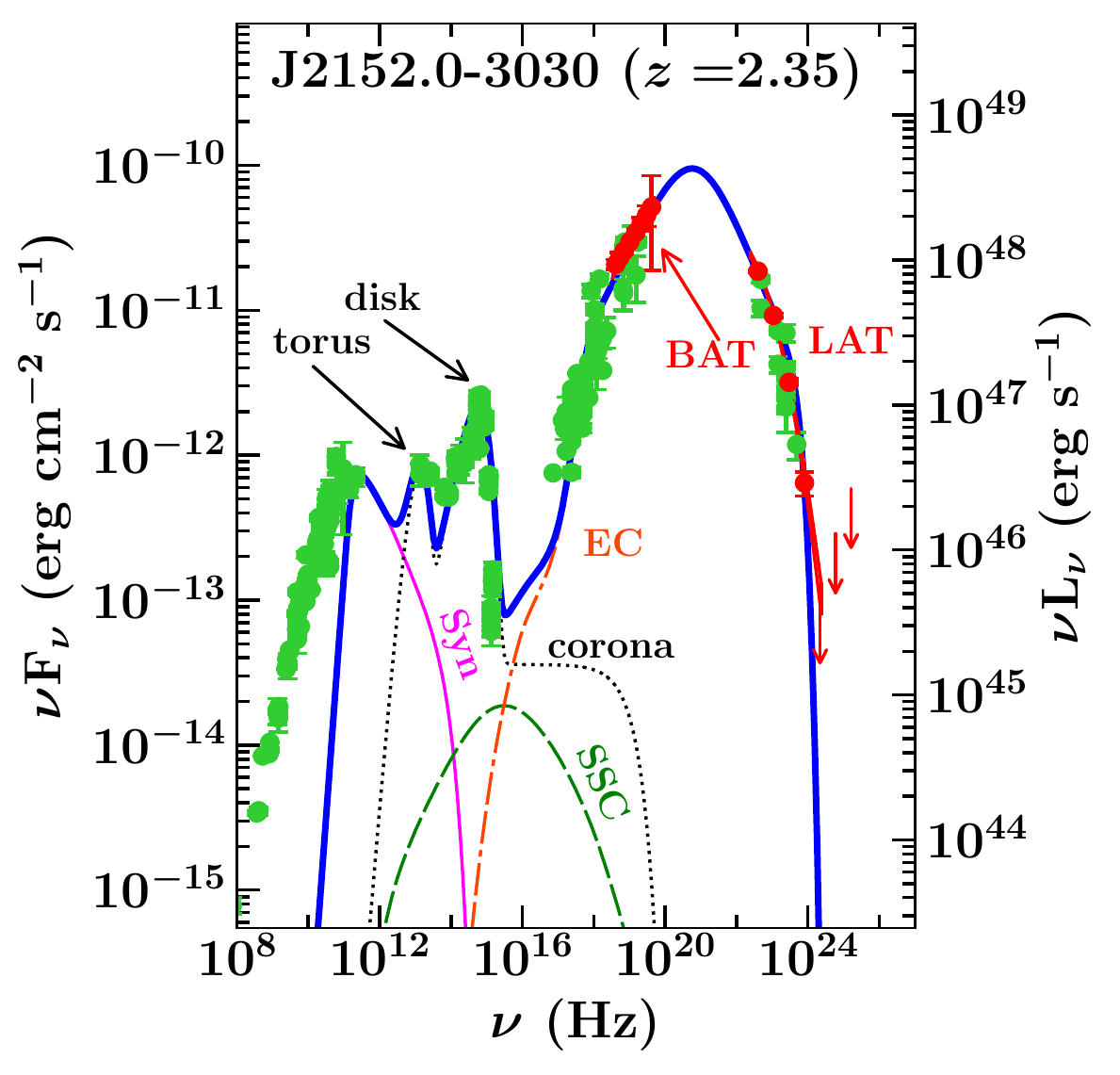}
\includegraphics[scale=0.7]{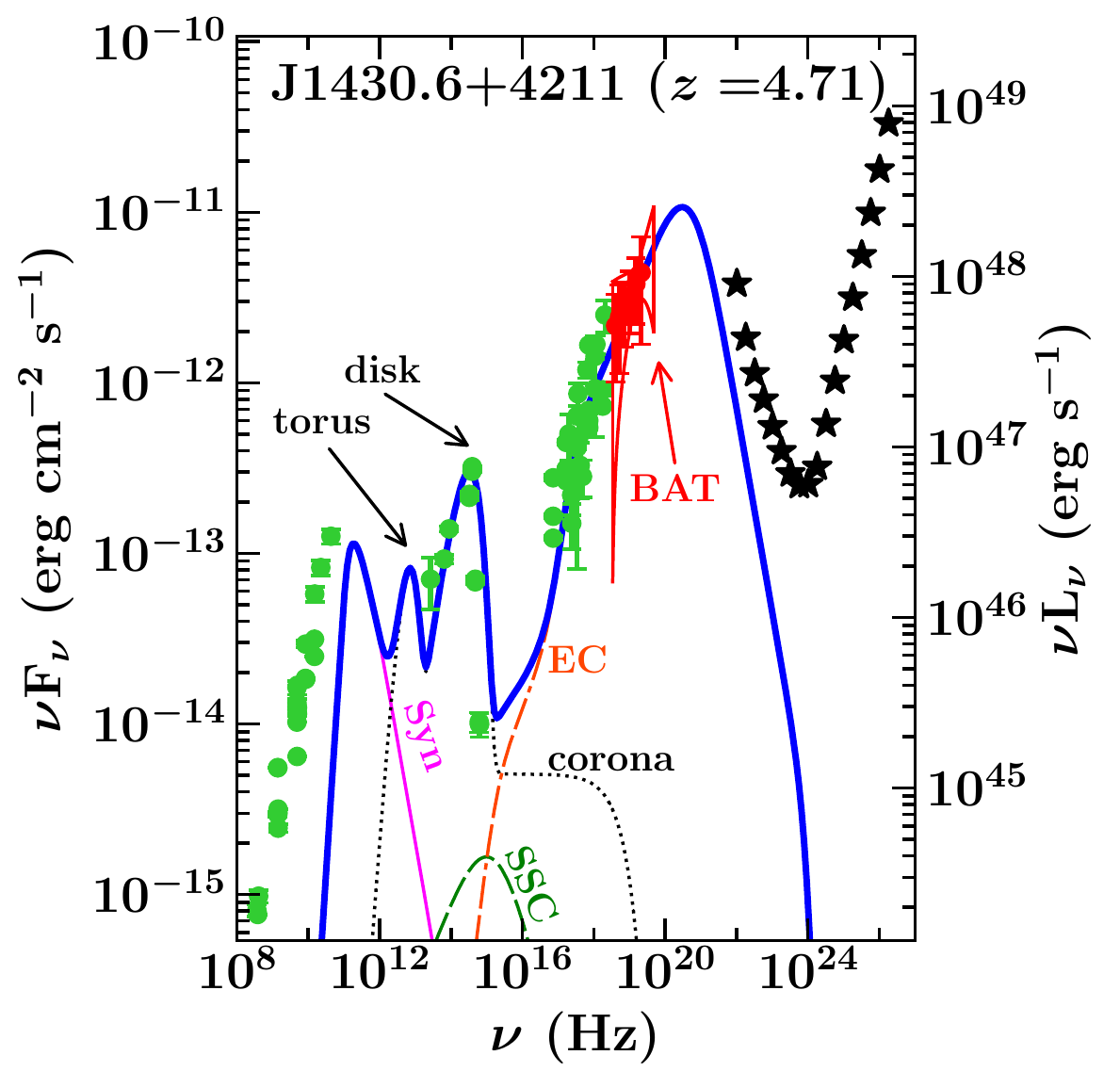}
}
\hbox{
\includegraphics[scale=0.7]{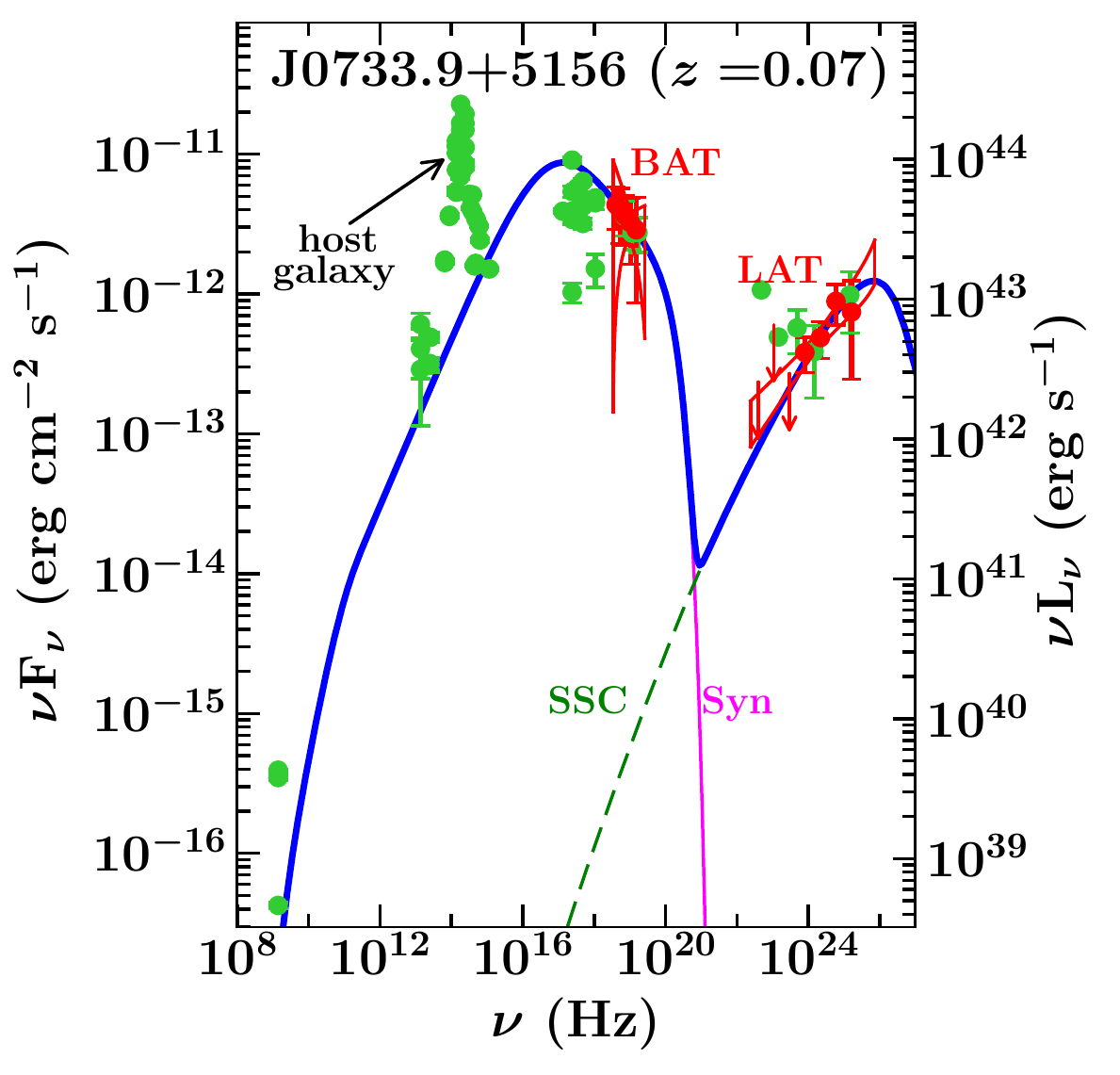}
\includegraphics[scale=0.7]{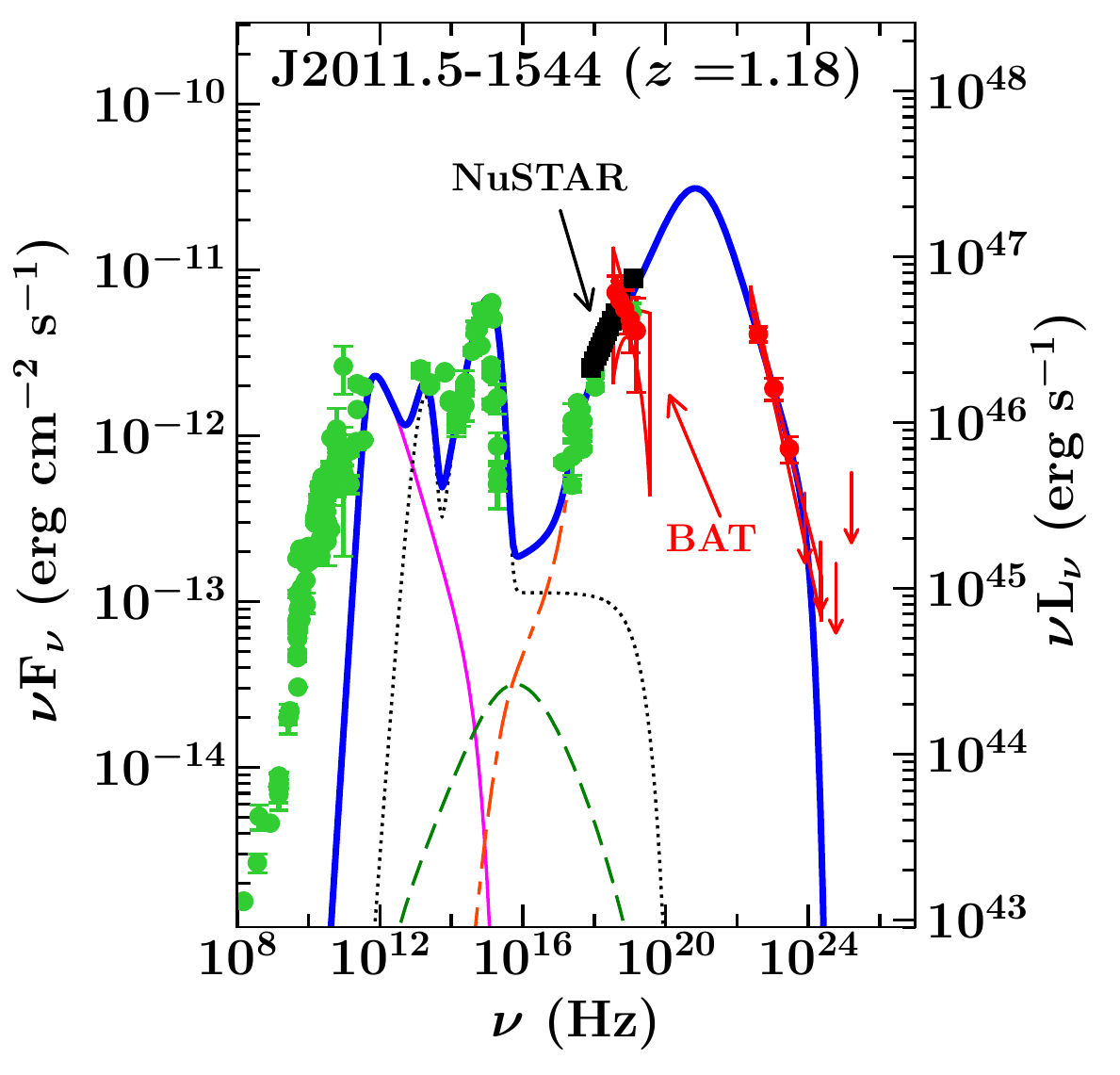}
}
\caption{Top: The modeled SED of the \fermi-LAT detected (left) and undetected (right) BAT blazars. \swift-BAT and \fermi-LAT spectral data are shown with red filled circles and bow-tie, while archival observations are displayed with lime-green filled circles. Thermal emissions from the torus, accretion disk, and the X-ray corona are represented with the black dotted line. We show the non-thermal synchrotron, SSC, and EC radiations with the pink thin solid,  green long-dashed and orange dash-dash-dot lines, respectively. The sum of all of the radiations is demonstrated with the blue thick solid line.  For the \gm-ray undetected sources, we plot 5$\sigma$ \fermi-LAT sensitivity for the duration covered in this work and toward the direction of the source with black stars. Bottom: The left panel illustrate the multi-wavelength SED of a BL Lac object. Note the IR-to-UV bump which is likely due to host galaxy emission. For sources whose BAT spectra do not match with the rest of their SED (right panel), we search for the availability of \nustar~data and use it for the modeling. The \nustar~data are shown with black squares. [{\it The modeled SED plots for all of the other blazars are shown in Figs. xxx$-$xxx in the electronic version.}]\label{fig:SED}}
\end{figure*}

For more than $>50$\% of sources, the near-infrared-to-ultraviolet (IR-to-UV) SED exhibits a bump which we interpret as due to accretion disk emission. By modeling this bump with a standard \citet[][]{1973A&A....24..337S} disk, we are able to constrain both the accretion disk luminosity ($L_{\rm d}$) and the mass of the central black hole ($M_{\rm bh}$) with the assumption of the accretion efficiency $\eta_{a}=0.1$ \citep[see also,][]{2002apa..book.....F,2010MNRAS.405..387G}.
In order to get independent measurements, we use results from our own ongoing optical spectroscopic campaign \citep[][Mejia-Restrepo et al. in prep.]{2017ApJ...850...74K} and also search for the availability of $M_{\rm bh}$ and $L_{\rm d}$ derived from single epoch optical spectroscopy available in the literature \citep[e.g.,][]{2012ApJ...748...49S}. We note that such commonly-used single epoch $M_{\rm bh}$ estimates carry systematic uncertainties of order $0.3-0.5$ dex \cite[see, e.g.,][for reviews of these methods]{2013BASI...41...61S,2014SSRv..183..253P}, and that the usage of the broad C\,\textsc{iv}\,$\lambda1549$ emission line may carry additional uncertainties related to the virialized nature of the BLR \cite[e.g.,][and references therein]{2012MNRAS.427.3081T,2018MNRAS.478.1929M}. Furthermore, if optical spectral parameters (e.g., line and continuum luminosities) are reported \citep[cf.][]{2012RMxAA..48....9T}, we use them to constrain $M_{\rm bh}$ following empirical relations commonly used for blazars \citep[e.g.,][]{2011ApJS..194...45S,2012ApJ...748...49S}. 
We compute the luminosity of the BLR from the emission line luminosities by considering the quasar emission line weights provided by \citet[][]{1991ApJ...373..465F}. Assuming that the BLR reprocesses 10\% of the disk radiation, we determine $L_{\rm d}$.

\subsection{Modeling constraints}
The SED model used in this work does not perform any statistical fitting and hence the uniqueness of the derived parameters cannot be claimed. However, our modeling efforts are driven by the quality of the multiwavelength data that provide fairly good constraints to the parameters. For example, we determine the $M_{\rm bh}$ and $L_{\rm d}$ either from the disk modeling or from the optical spectroscopy and these two parameters regulate the size and radiative energy densities of various external photon fields. The shape of the X-ray spectrum constrains the relative dominance of the EC and SSC mechanisms since different parts of the electron energy distribution are involved. The SSC radiation regulates the size of the emission region and also the magnetic field. A Compton dominated SED reflects the prevalence of the external photon densities over the magnetic one and since in our model these are a function of the distance of the emission region from the central black hole, we are able to constrain $R_{\rm d}$. In Compton-dominated blazars, the low- and high-energy indices of the broken power law electron energy distribution are determined from the spectral shapes of the X-ray and \gm-ray SEDs, respectively. The optical-UV spectrum further controls the high-energy index provided it is dominated by the falling synchrotron emission. On the other hand, in blazars whose synchrotron peak lies at UV-X-ray energies, i.e., HSP sources, we use \fermi-LAT spectra to determine the low-energy index. The shape of the BAT spectrum, along with the soft X-ray SED, provides a good constraint to the high-energy index in such objects.

\section{Physical Characteristics}\label{sec5}
\begin{figure*}
\gridline{\fig{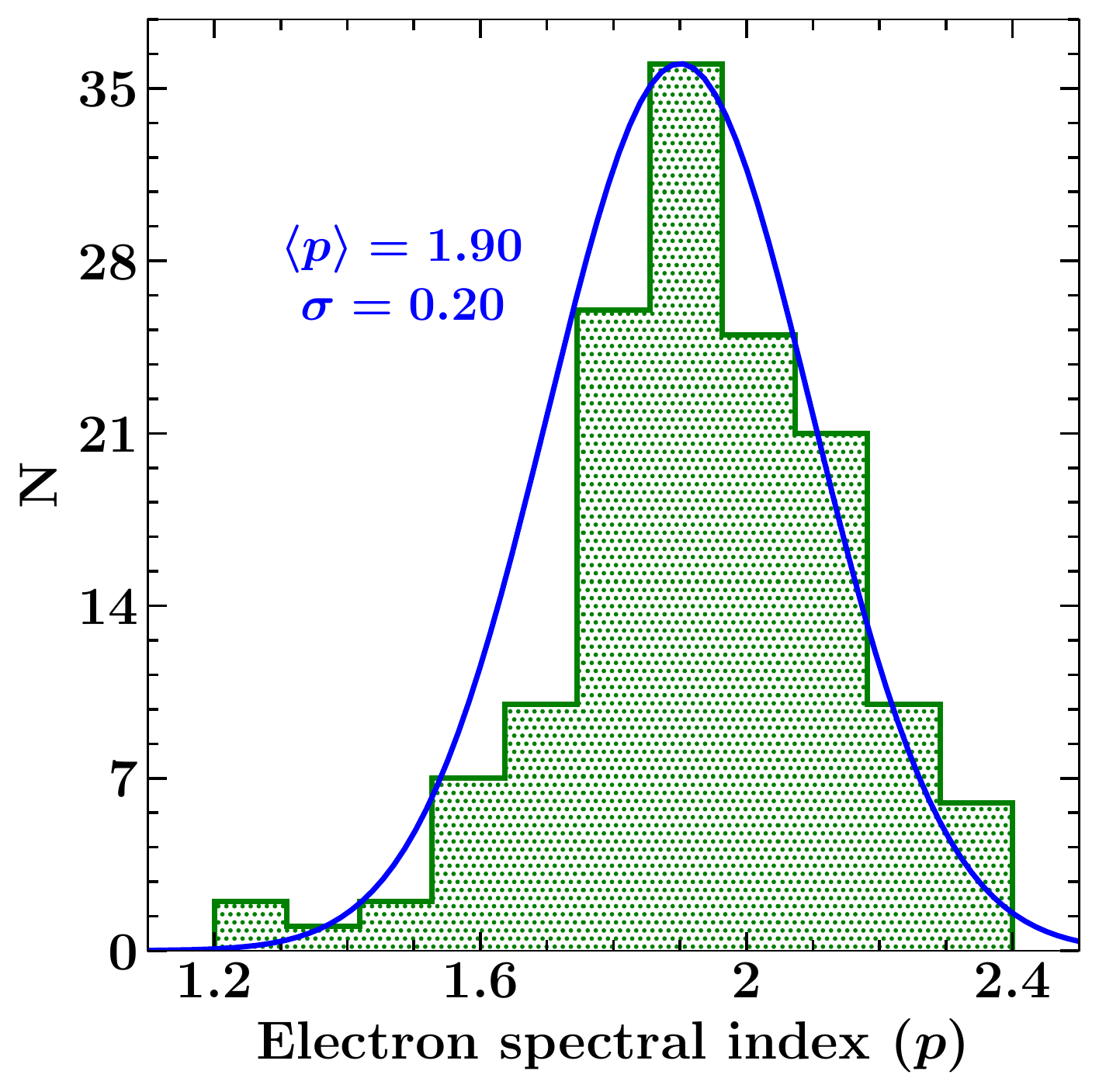}{0.24\textwidth}{(a)}
          \fig{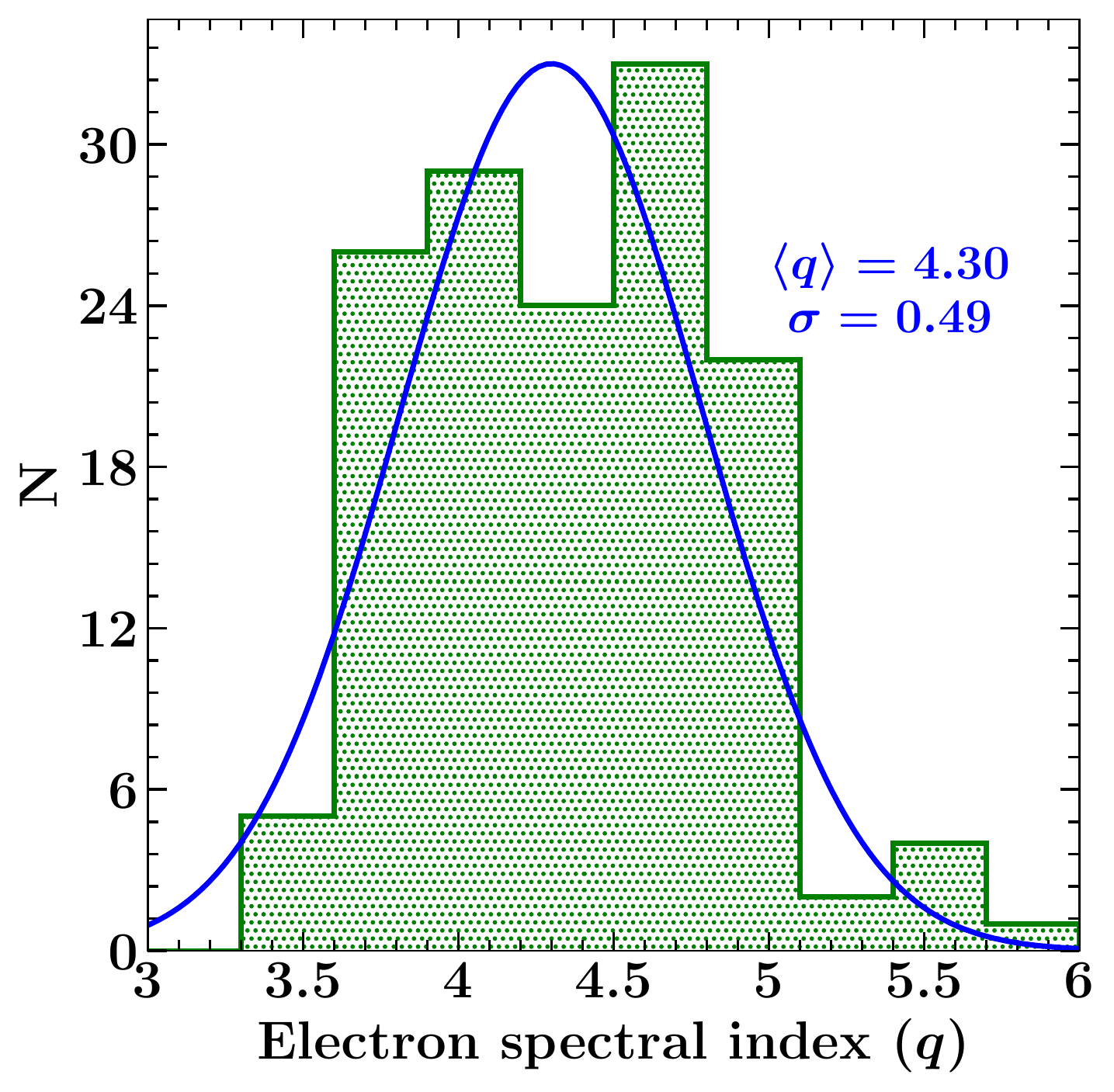}{0.24\textwidth}{(b)}
          \fig{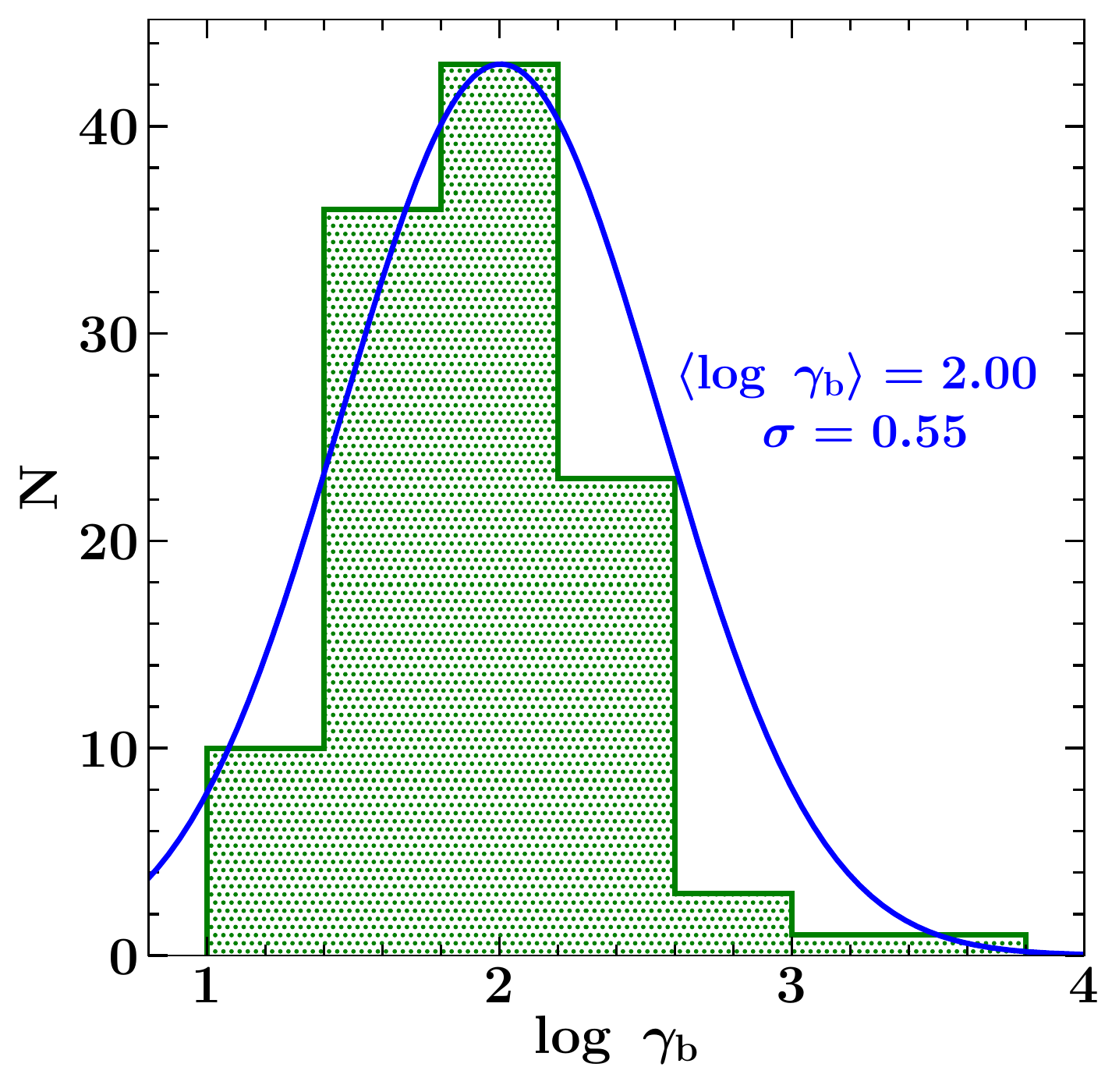}{0.24\textwidth}{(c)}
          \fig{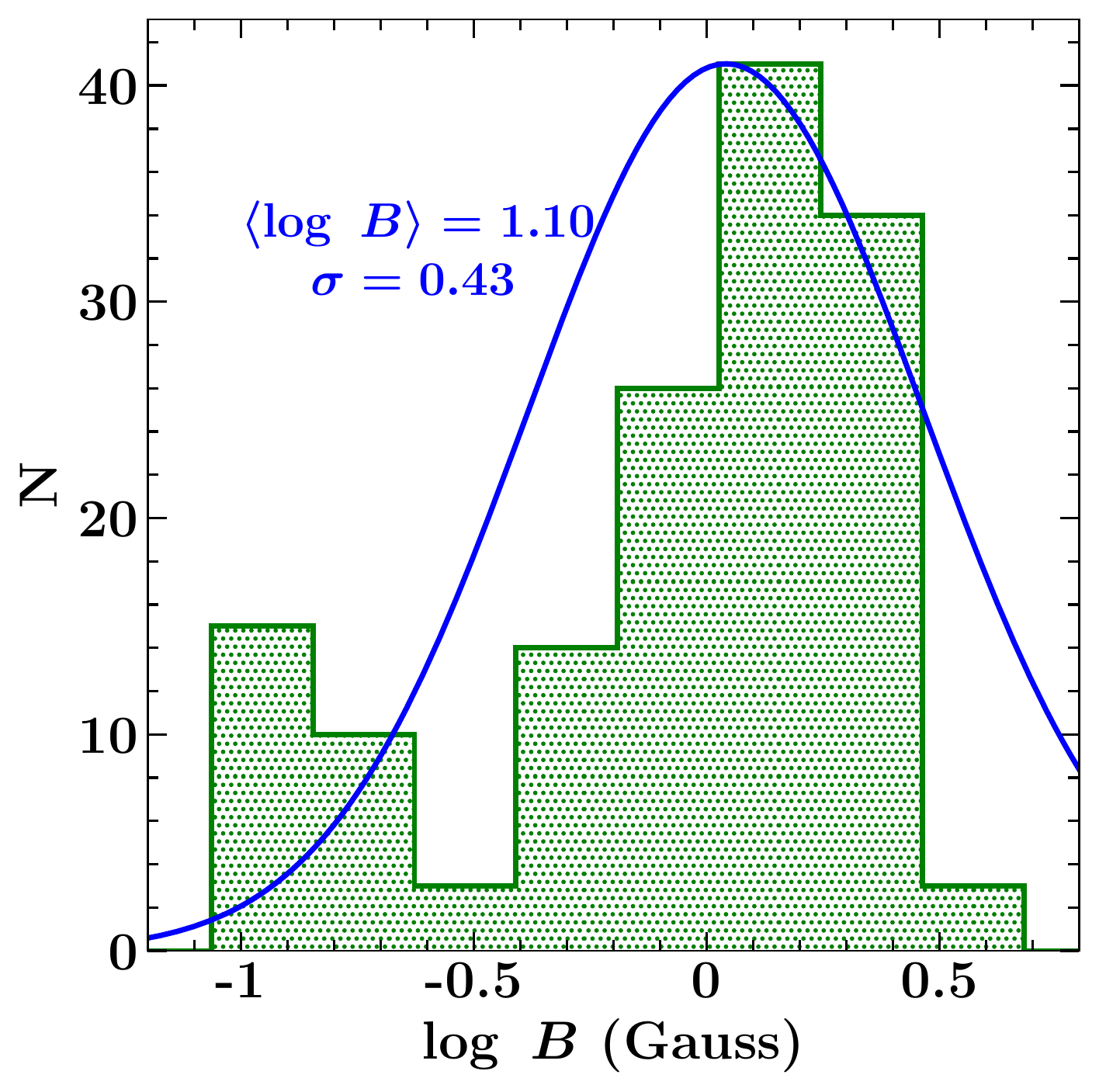}{0.24\textwidth}{(d)}
          }
\gridline{\fig{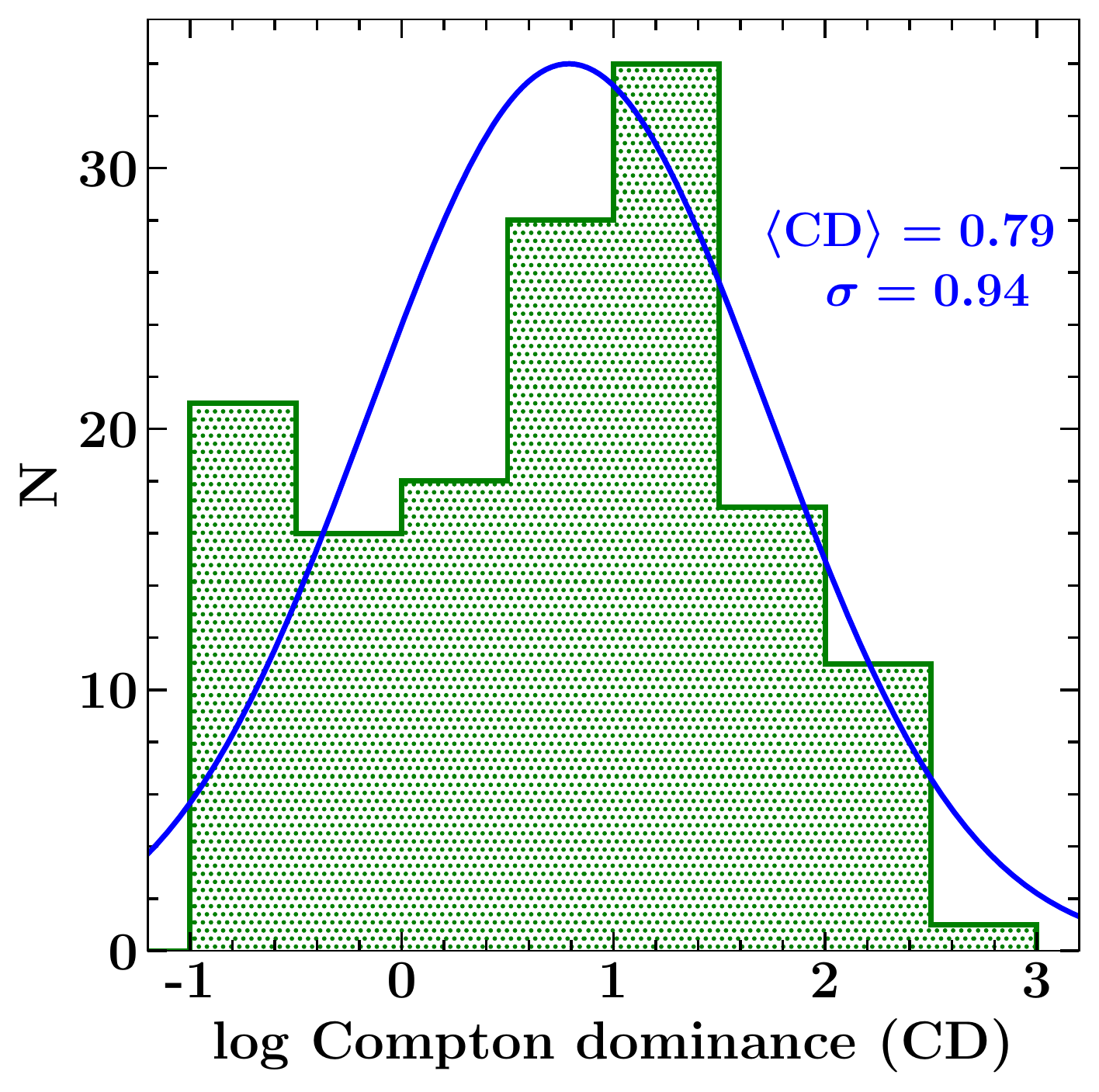}{0.24\textwidth}{(e)}
          \fig{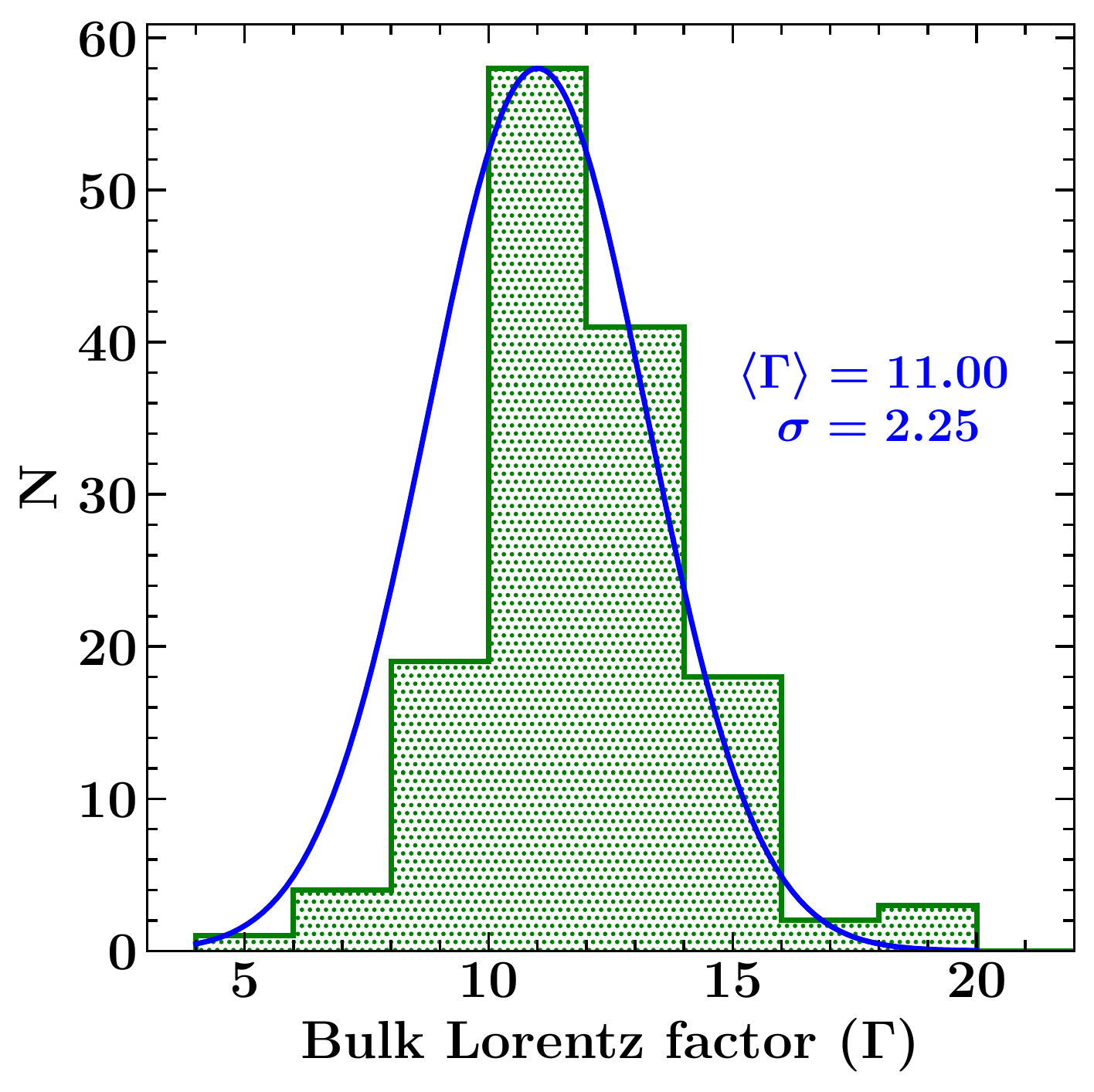}{0.24\textwidth}{(f)}
          \fig{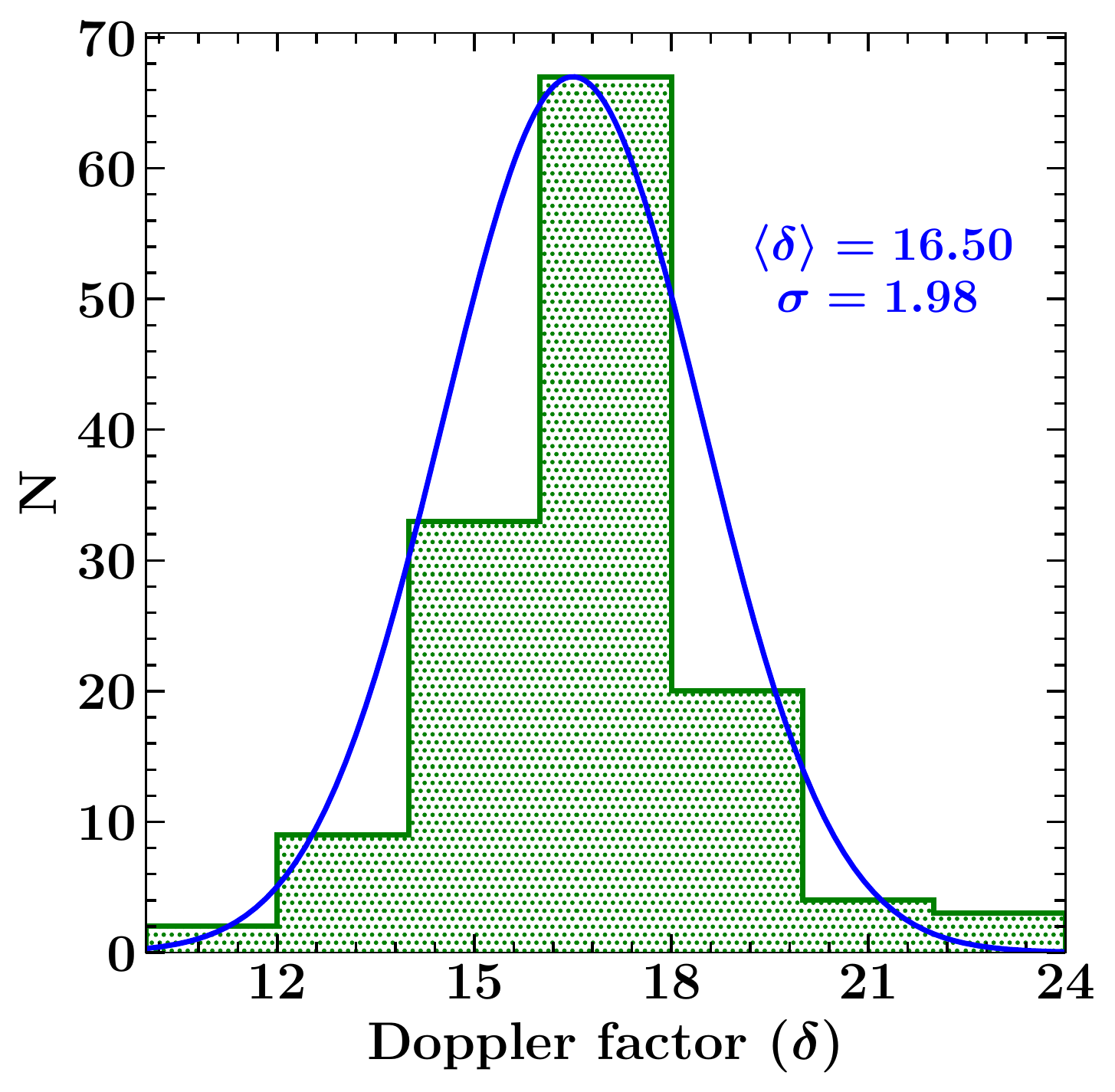}{0.24\textwidth}{(g)}
          \fig{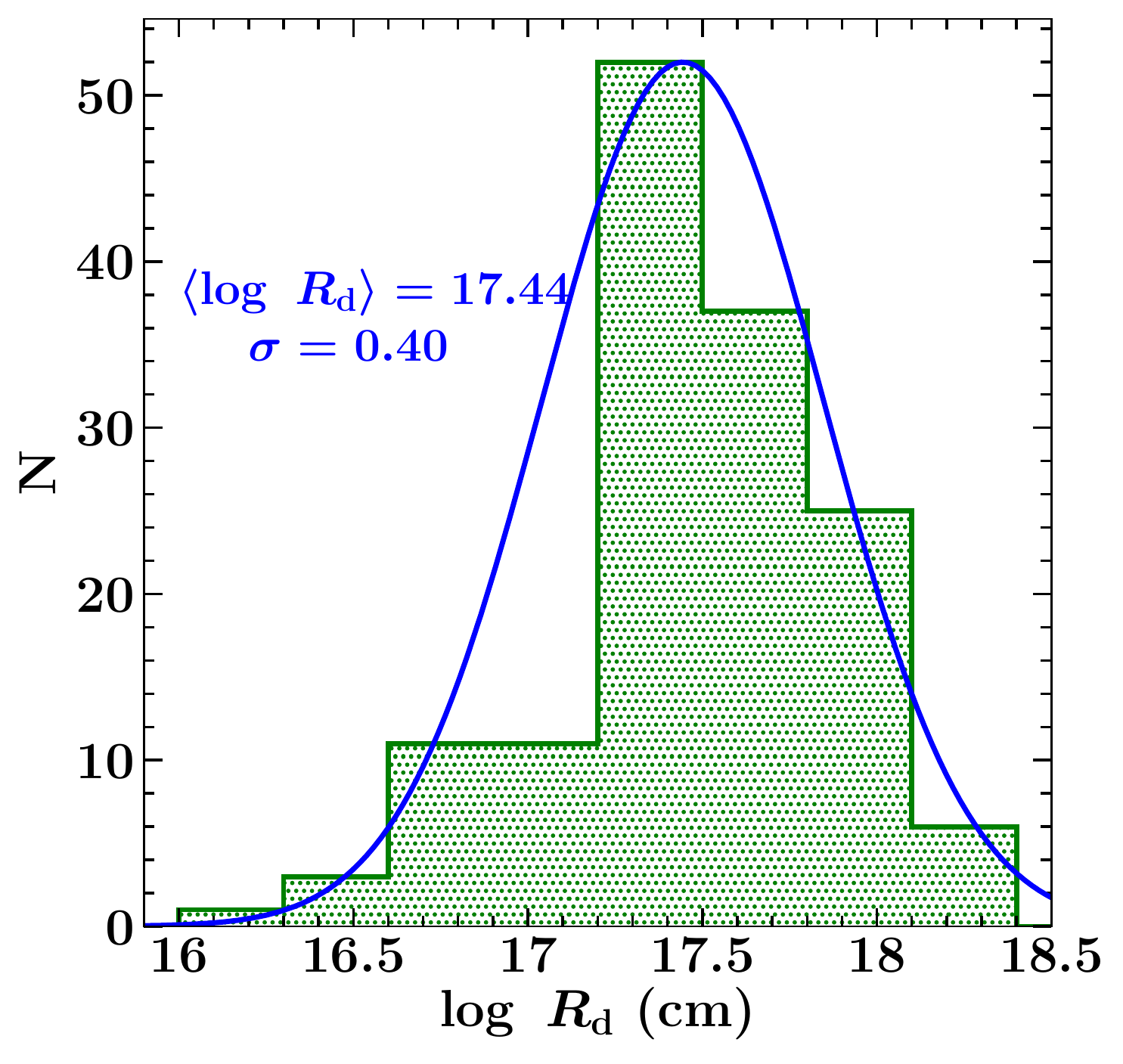}{0.24\textwidth}{(h)}
          }
\caption{Histograms of modeled SED parameters for the 142 blazars in our sample. We fit the data with a Gaussian distribution, shown as a blue line, and the standard deviations are quoted.\label{fig:histo}}
\end{figure*}

In the left panel of Figure~\ref{fig:lumin}, we show the 14$-$195 keV rest-frame luminosity of various types of BAT detected AGNs as a function of their redshifts. As can be seen that blazars dominate the 105-month BAT catalog above $z\approx0.5$. This is likely a selection effect due to relativistic beaming which makes blazar jets brighter compared to other class of astrophysical objects. Considering only blazars, the photon index versus luminosity distribution is shown in the middle panel and we highlight $z>2$ blazars by showing them with pink stars. As can be seen, these are the most powerful ones with luminosity exceeding $10^{47}$ \lum. Furthermore, there is a hint of an anti-correlation (Spearmann coefficient $\rho= -0.29\pm0.08$, probability of no correlation or PNC = 0.01) where more luminous objects, mostly FSRQs, tend to have a harder BAT spectrum. This is because the X-ray to \gm-ray emission in FSRQs primarily originates via EC process which has a hard spectral shape in X-rays (see Figure~\ref{fig:SED}) which becomes more prominent at high redshifts due to shift of SED peaks to lower frequencies and $K$-correction. Less luminous HSP blazars generally have a falling synchrotron emission in the BAT energy range and thus exhibit a steep spectrum with 14$-$195 keV photon index $>$2. We plot \gm-ray luminosity as a function of the hard X-ray one in the right panel of Figure~\ref{fig:lumin}. Interestingly, the \fermi-LAT undetected objects share the same range of the 14$-$195 keV luminosity with \gm-ray detected ones. Since all of them are FSRQs, this observation indicates their SED peaks to lie at lower frequencies with respect to \fermi-LAT sources.

The modeled SEDs of BAT blazars are shown in Figure~\ref{fig:SED} and we provide associated parameters in Table~\ref{tab:sed_param}. Note that there are 28 sources whose broadband SEDs are well represented with synchrotron-SSC processes, i.e., without invoking EC mechanism. All of them are HSP blazars and their names and redshifts are provided in Table~\ref{tab:syn}. 

{\it Particle energy distribution:} The histograms of the spectral indices of the electron energy distribution are shown in Figure~\ref{fig:histo} (panels a and b). The low-energy index ($p$) distribution peaks at 1.90 and when fitted with a Gaussian function, the dispersion is $\sigma=0.20$. On the other hand, the high-energy index ($q$) has a rather broad range. The distribution peaks at $\langle q \rangle=4.30$ with $\sigma=0.49$ and its tail extends up to $q\sim6$. This is primarily due to the inclusion of \gm-ray undetected blazars whose falling part of the inverse Compton spectrum needs to be steep to remain below the \fermi-LAT detection threshold. Additionally, the steep index can also be due to a few HSP blazars with falling BAT spectrum. The distribution of the break energy $\gamma_{\rm b}$, which is derived from the synchrotron peak frequency location, peaks at 100 and tails to very large values, $\gtrsim$1000. Very large $\gamma_{\rm b}$ is probably due to inefficient cooling of relativistic electrons in HSP blazars and overall the distribution has a broad width ($\sigma=0.55$, Figure~\ref{fig:histo}, panel c).

{\it Magnetic field:} The magnetic field and Compton dominance\footnote{Compton dominance or CD is defined as the ratio of the inverse Compton to synchrotron peak luminosities.} distributions are shown in panels (d) and (e) of Figure~\ref{fig:histo}. There is a hint of bi-modality that can be understood keeping in mind that a major fraction of BAT blazars studied here are FSRQs and remaining are mostly HSP blazars with synchrotron dominated SEDs. FSRQs are known to have Compton dominated (CD$>$1) SEDs with a larger magnetic field compared to HSP blazars \citep[see, e.g.,][]{2010MNRAS.401.1570T,2017ApJ...851...33P}. Therefore, the distributions are skewed towards larger CD and $B$ with $\langle \log{\rm CD} \rangle=0.79$ and $\langle B \rangle=1.10$ Gauss, respectively.

{\it Bulk Lorentz factor and Doppler factor:} In panels (f) and (g) of Figure~\ref{fig:histo}, we show the histograms of the bulk Lorentz factor ($\Gamma$) and Doppler factor ($\delta$). Both of them have narrow distributions peaking at $\langle \Gamma \rangle=11$ and $\langle \delta \rangle=16.5$. These values are in agreement with that found from radio studies \citep[e.g.,][]{2010A&A...512A..24S}, from the SED modeling of a large sample of \gm-ray emitting blazars \citep[][]{2017ApJ...851...33P}, and also from luminosity function studies \citep[][]{2012ApJ...751..108A}.

 \begin{figure}
\hbox{\hspace{0cm}
\includegraphics[scale=0.5]{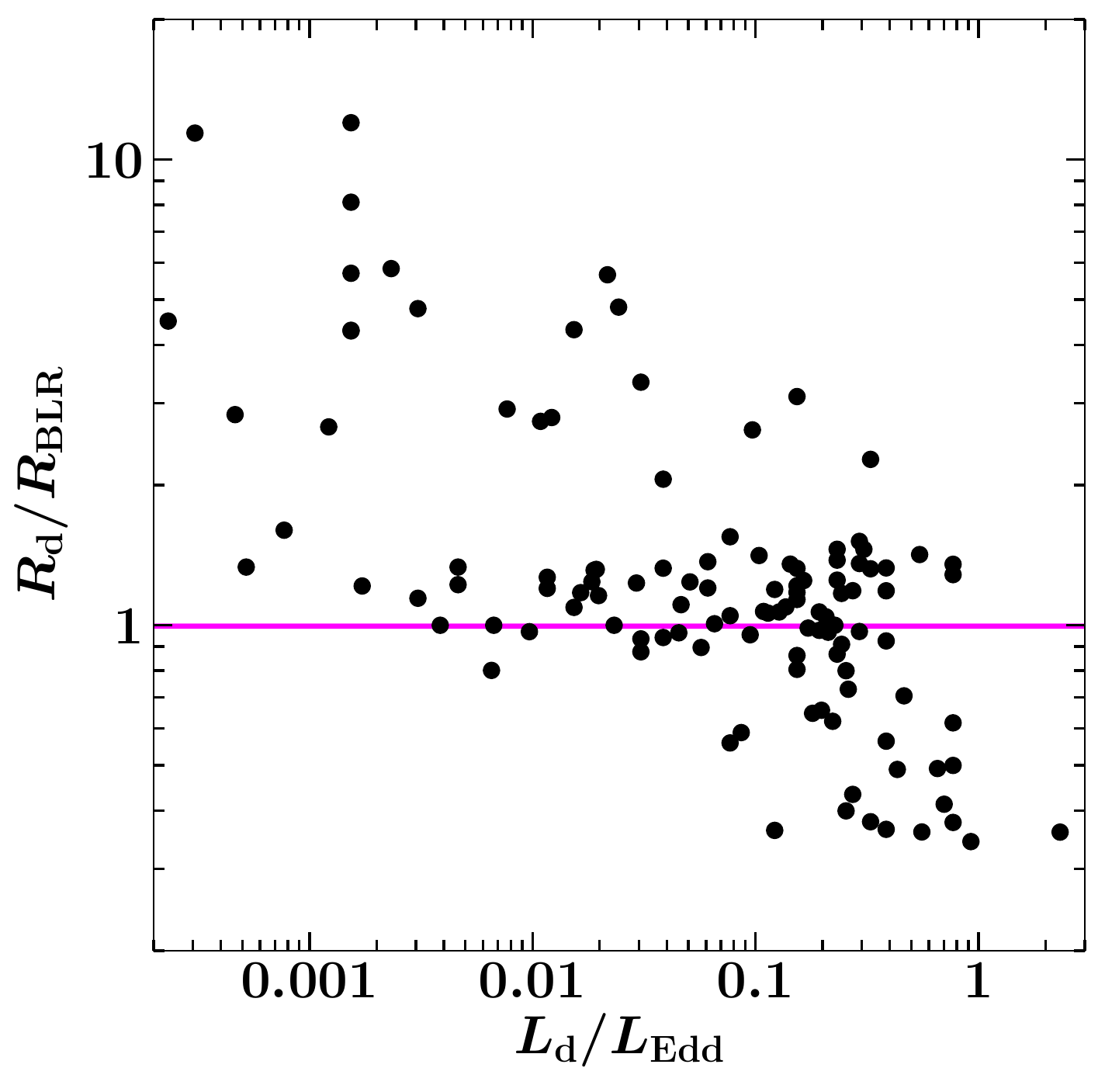}
}
\caption{The $R_{\rm d}$ as a function of the accretion disk luminosity $L_{\rm d}$. The horizontal line represents the inner boundary of the BLR.} \label{fig:blr}
\end{figure}

{\it Dissipation distance:} The distribution of the location of the emission region peaks at $R_{\rm d}\sim3\times10^{17}$ cm with a majority of sources having $R_{\rm d}>10^{17}$ cm (see Figure~\ref{fig:histo}, panel h). To understand the emission region location with respect to BLR, we plot $R_{\rm d}/R_{\rm BLR}$ as a function of $L_{\rm d}$ in Eddington units in Figure~\ref{fig:blr}. 
The $R_{\rm BLR}$ is estimated following \citet[][]{2009MNRAS.397..985G}, i.e.,  $R_{\rm BLR}=10^{17}\,(L_{\rm d}/10^{45}\,{\rm erg\,s}^{-1})^{1/2}$ cm. The location is outside BLR, i.e., above the horizontal line, mostly for blazars with less luminous accretion disks which is typically the case of BL Lacs. Moreover, a majority of sources have $R_{\rm d}$ comparable to or slightly larger than $R_{\rm BLR}$, thus indicating the jet environment surrounding the emission region to be transparent enough for \gm-rays to avoid absorption via pair production with BLR photons \citep[see also,][for quantitative discussions]{2016ApJ...821..102B,2018ApJ...863...98P}. Blazars with most luminous accretion disks, i.e. with the largest BLR, tend to have emission region well within it. These are the objects with the largest Compton dominance.

 \begin{figure*}[t!]
\hbox{\hspace{0cm}
\includegraphics[scale=0.4]{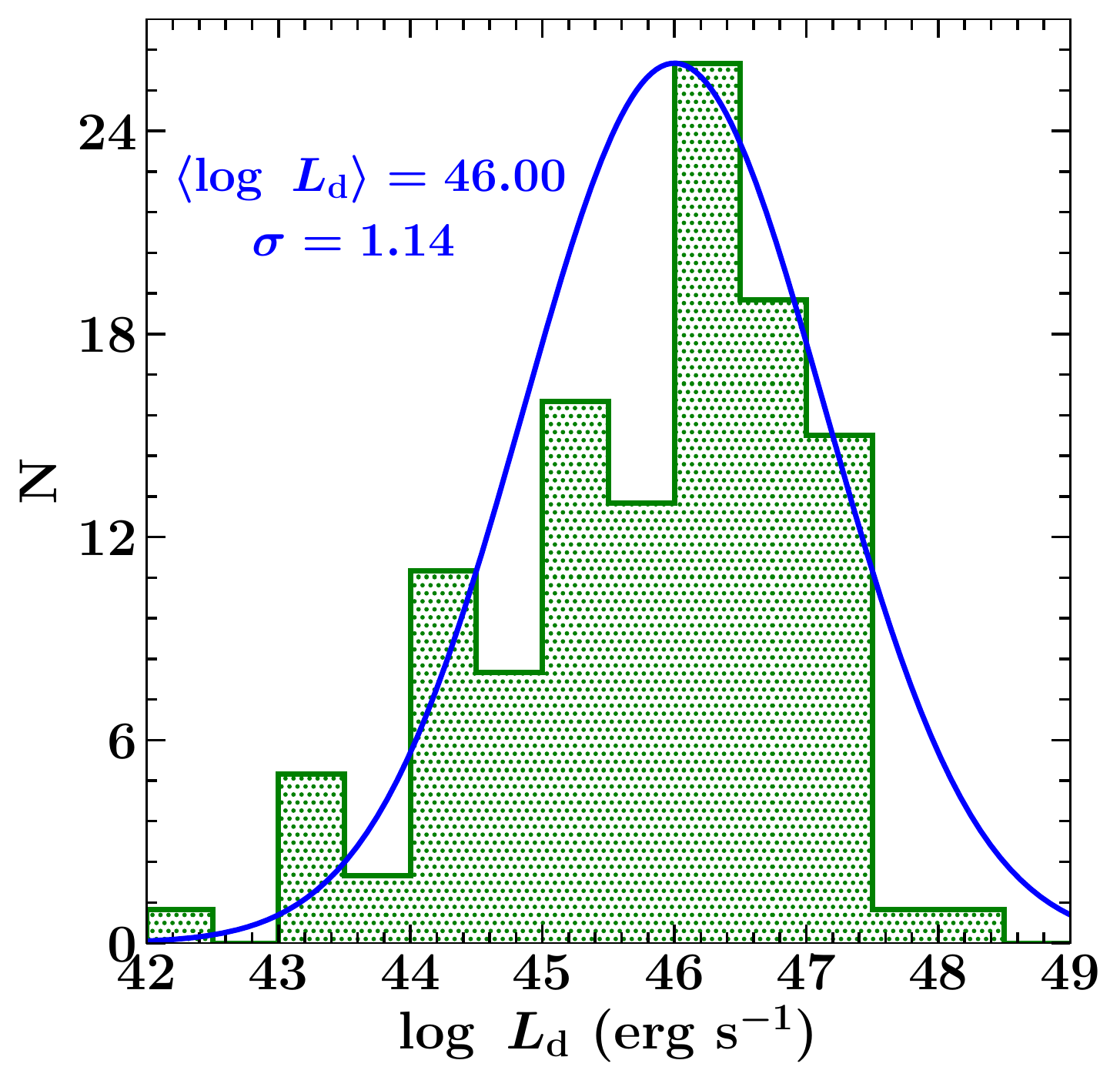}
\includegraphics[scale=0.4]{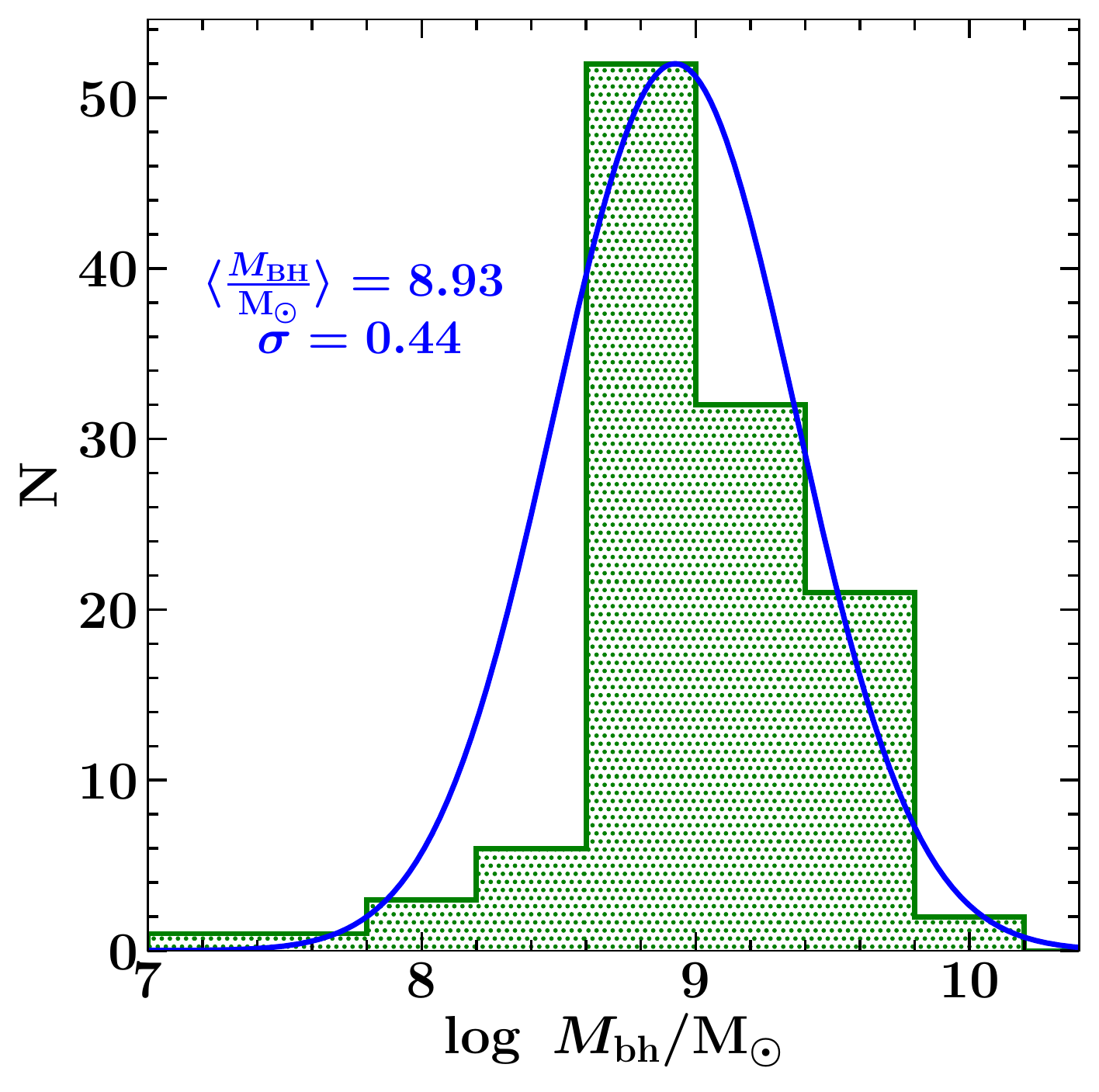}
\includegraphics[scale=0.4]{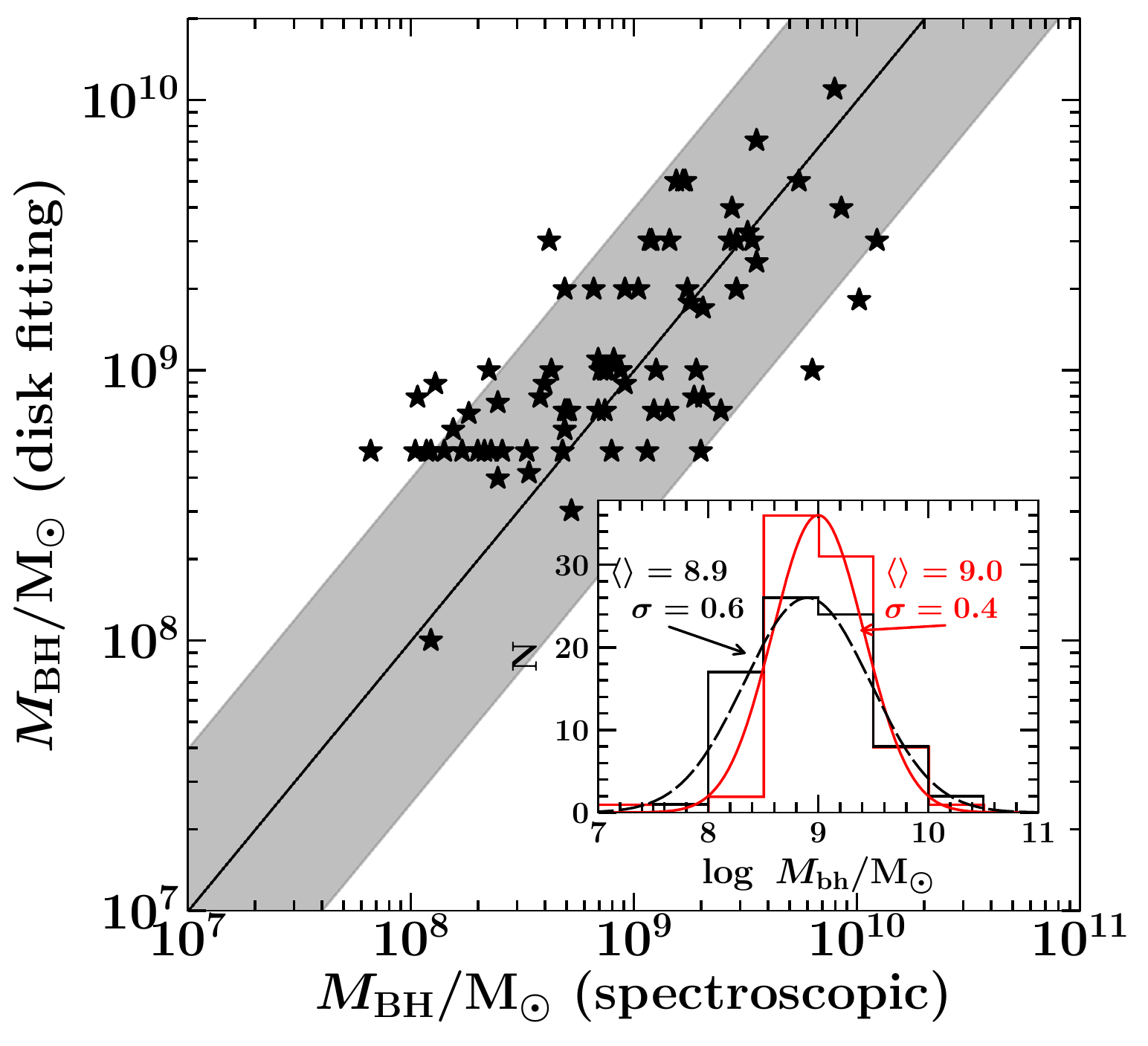}
}
\caption{The distributions of $L_{\rm d}$ (left) and $M_{\rm bh}$ (middle) for BAT blazars studied in this work. In the right panel, we compare $M_{\rm bh}$ values derived from the optical spectroscopy and disk modeling methods. As can be seen, both approaches reasonably agree with each other. Inset shows the histograms of the plotted quantities. The shaded area represent the uncertainty of a factor of 4 associated with the virial estimation \citep[e.g.,][]{2011ApJS..194...45S}.} \label{fig:mbh}
\end{figure*}

 \begin{figure*}[t!]
\hbox{\hspace{0cm}
\includegraphics[scale=0.6]{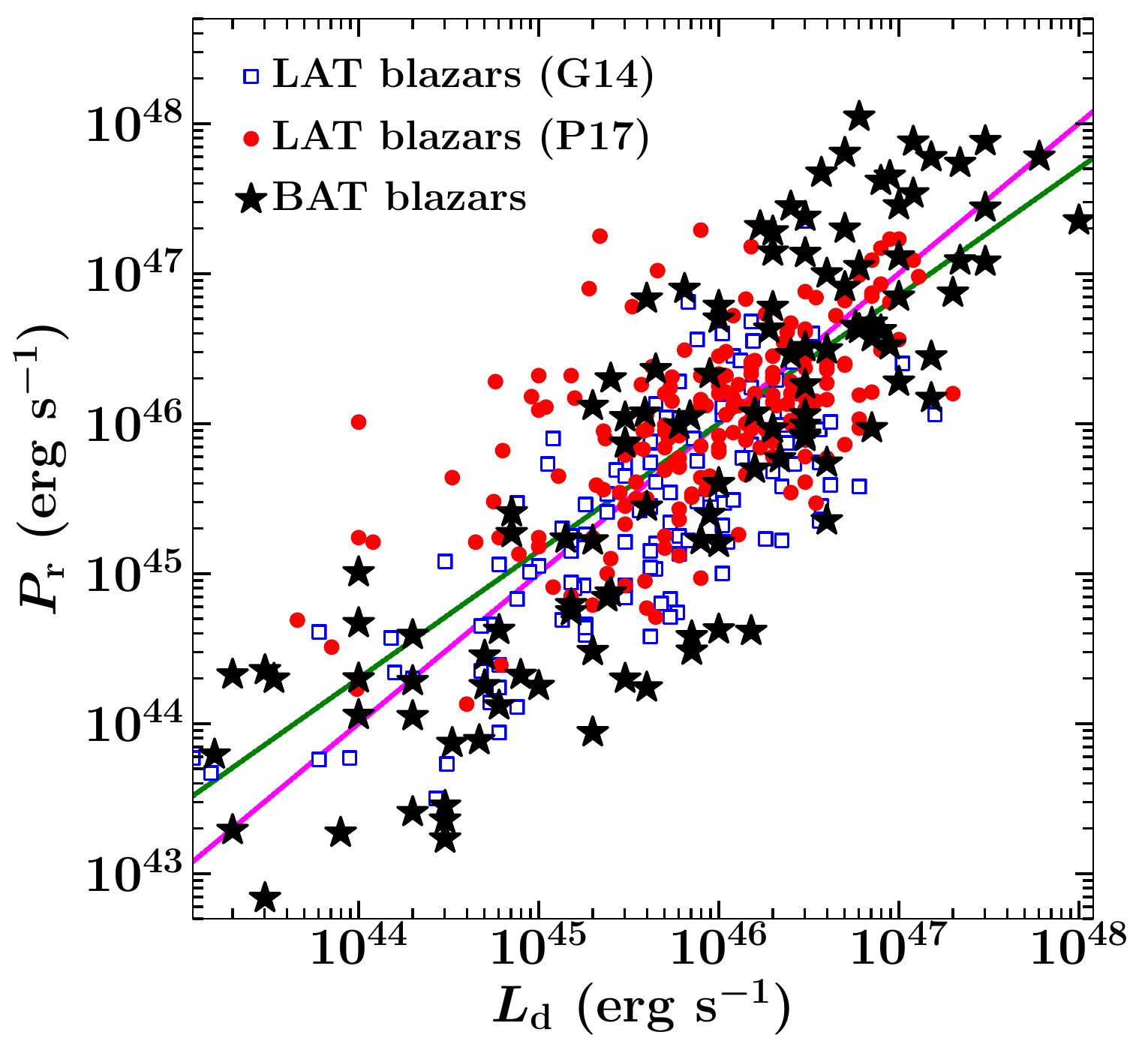}
\includegraphics[scale=0.6]{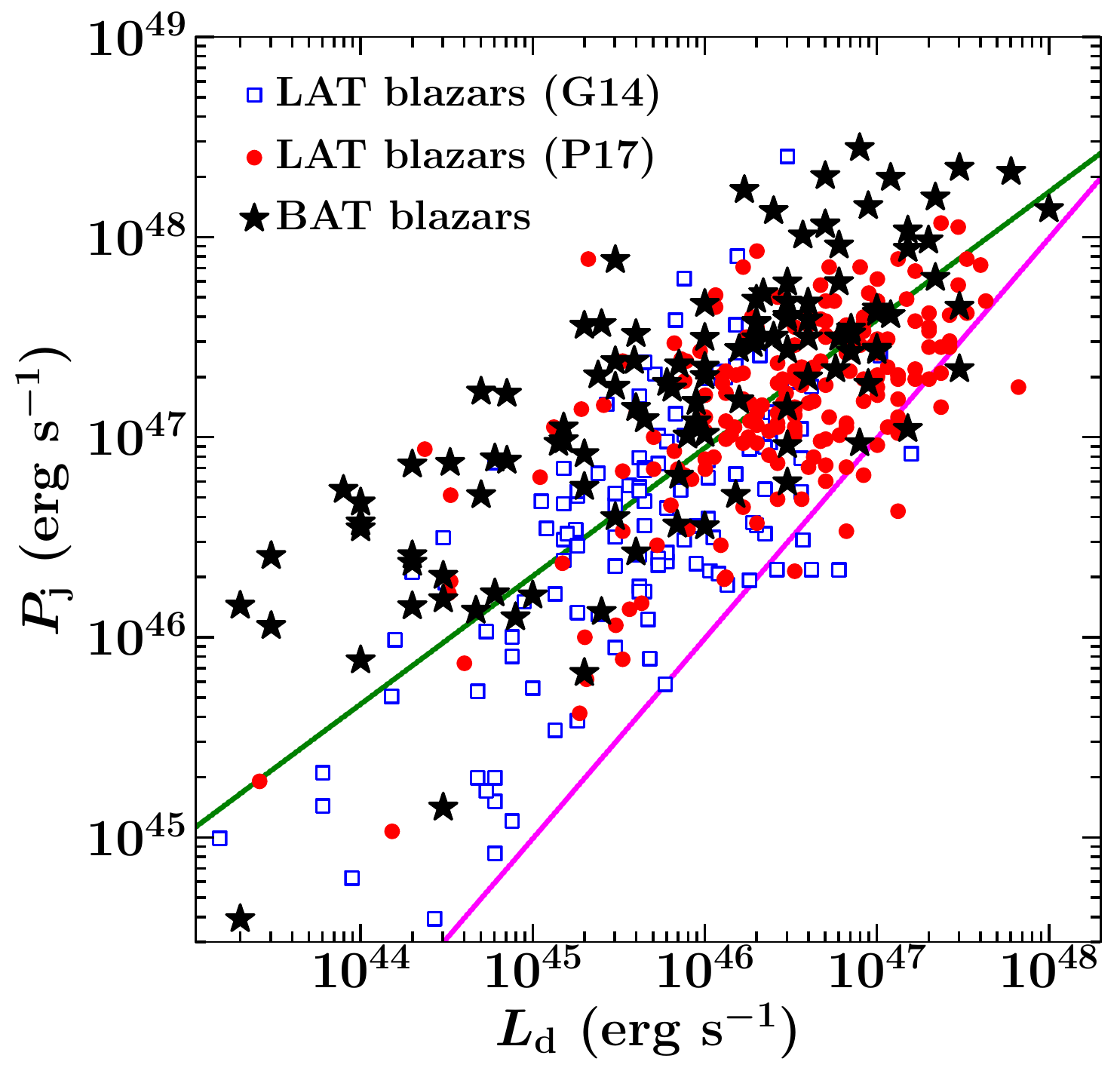}
}
\caption{Left: The power that jet produces in the form of radiation ($P_{\rm r}$) as a function of the luminosity of the accretion disk. Right: The total jet power ($P_{\rm j}=P_{\rm p}+P_{\rm e}+P_{\rm m}$) as a function of the $L_{\rm d}$. In both the plots, \fermi-LAT blazars are displayed with red circles \citep[][P17]{2017ApJ...851...33P} and empty blue squares \citep[][G14]{2014Natur.515..376G}, whereas, BAT blazars are shown with black stars. The pink and green solid lines refer to the equality and best-fit of the plotted quantities, respectively.} \label{fig:bat-lat}
\end{figure*}

{\it Central engine:} We have derived $M_{\rm bh}$ in 92 objects using the disk modeling approach which show the signature of the accretion disk emission at optical-UV energies. In 20 sources, we use the optical spectroscopic $M_{\rm bh}$ since their optical-UV SED is synchrotron dominated. The multiwavelength SEDs of 28 HSP blazars (Table~\ref{tab:syn}) are well reproduced with a synchrotron-SSC model and thus no $M_{\rm bh}$ value was estimated/used. Remaining 6 sources neither exhibit the big blue bump nor have the spectroscopic $M_{\rm bh}$. In these objects, we suitably assume a $M_{\rm bh}$ value based on the available observations. The different methods adopted to compute $M_{\rm bh}$ are tabulated in Table~\ref{tab:sed_param}.

The left and middle panels of Figure~\ref{fig:mbh} represent the $M_{\rm bh}$ and $L_{\rm d}$ histograms for BAT blazars. The $M_{\rm bh}$ distribution has a rather narrow range ($\sim10^8-10^{10}$ M$_{\odot}$) and peaks at $\langle \log M_{\rm bh} \rangle=9.0$ M$_{\odot}$. Fitting it with a log-normal function returns a width of $\sigma=0.44$. On the other hand, $L_{\rm d}$ peaks at 10$^{46}$ \lum~and has a rather broad range with $\sigma=1.14$. 

In our sample, there are a total of 82 blazars that have $M_{\rm bh}$ measurements from both the disk-fitting and optical spectroscopy methods. In the right panel of Figure~\ref{fig:mbh}, we compare $M_{\rm bh}$ derived using these two approaches. As can be seen, both methods provide reasonably similar $M_{\rm bh}$ for BAT blazars and thus confirm the findings earlier reported for \fermi-LAT detected sources \citep[][]{2017ApJ...851...33P}.

 \begin{figure*}
\hbox{\hspace{0cm}
\includegraphics[scale=0.6]{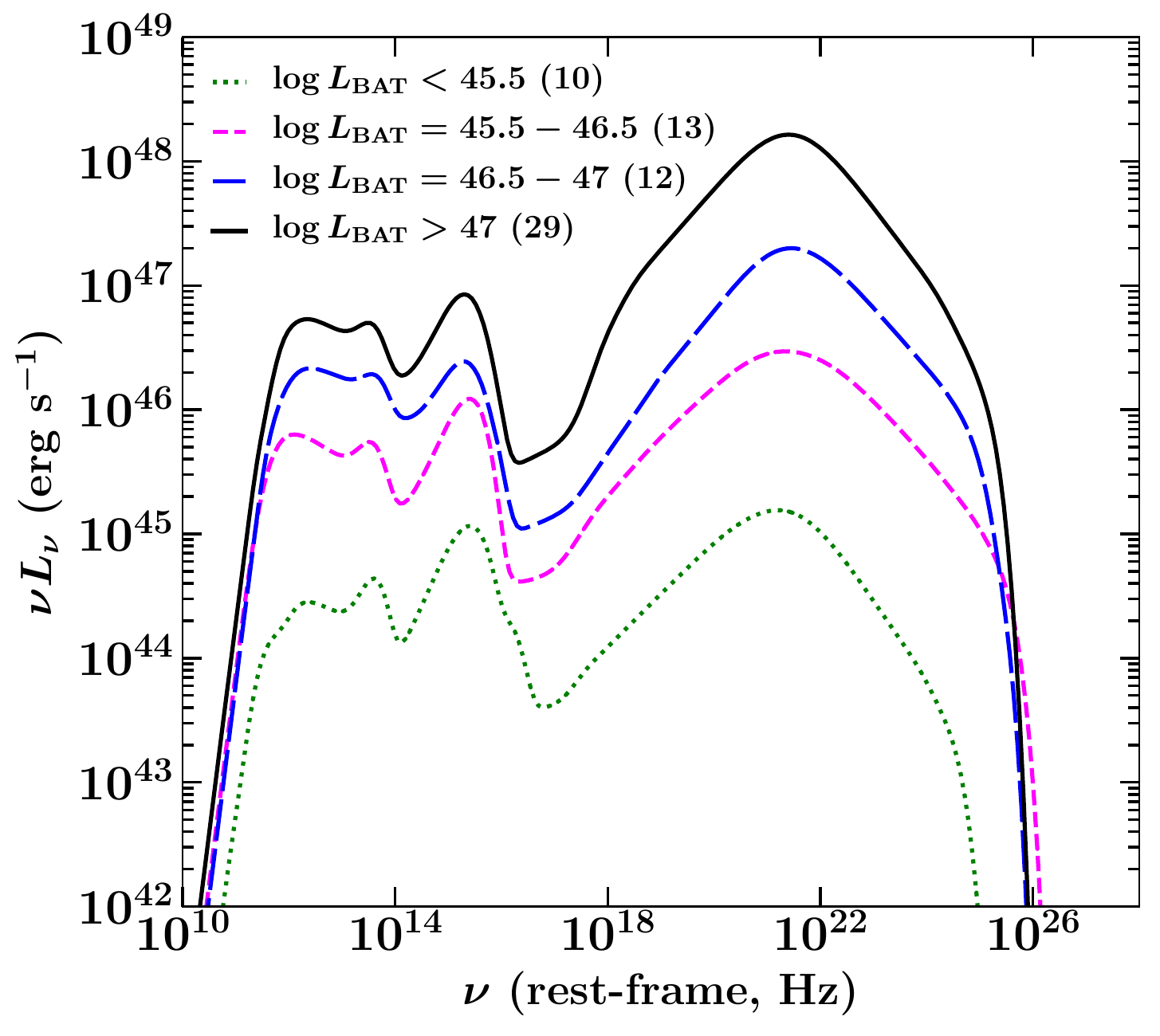}
\includegraphics[scale=0.6]{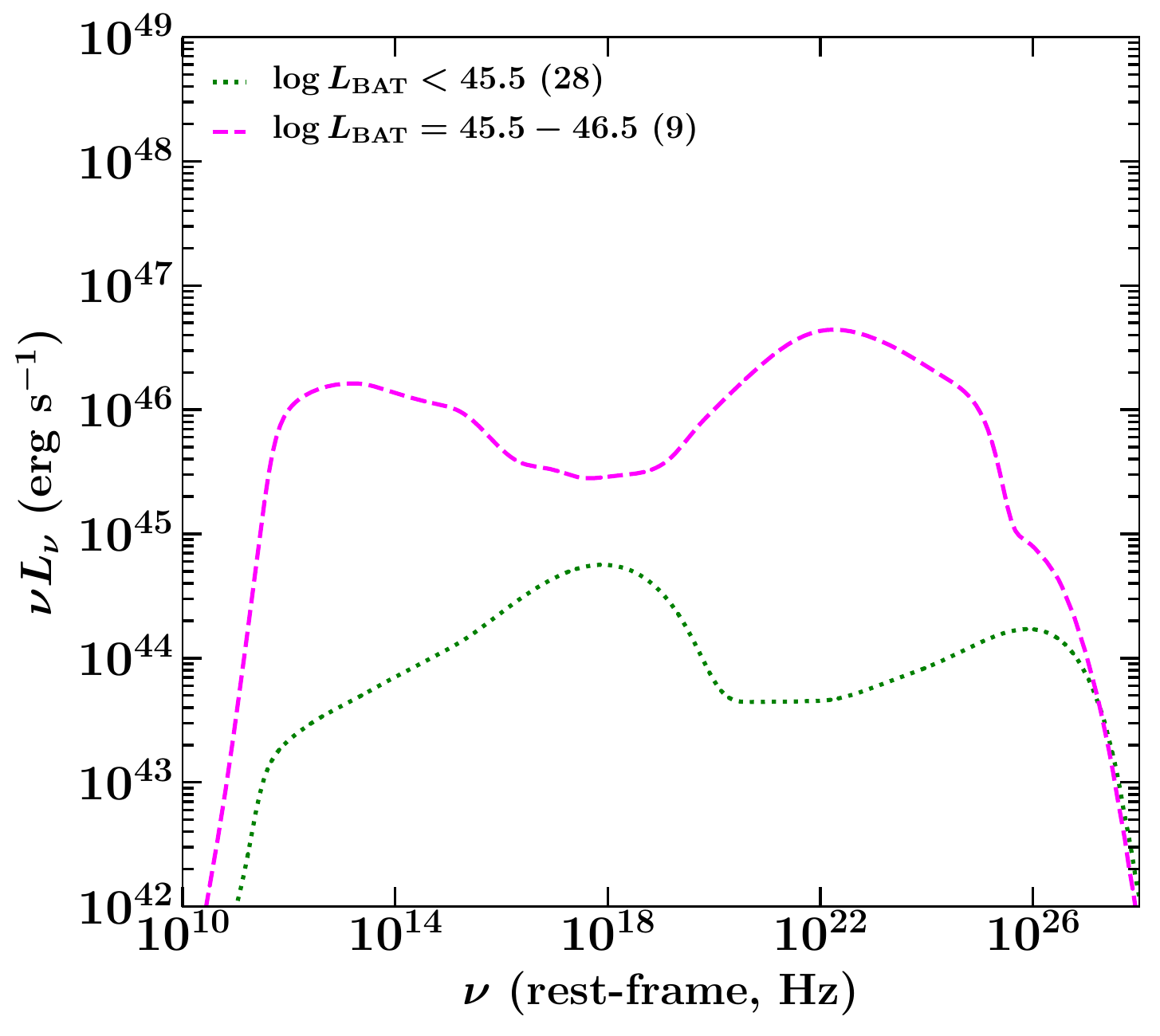}
}
\caption{Left: Averaged SEDs of BAT FSRQs in four different $L_{\rm BAT}$ bins, as labeled. Right: Same as left but for BL Lacs. Note that there are no BL Lac objects detected with \swift-BAT with $L_{\rm BAT}>10^{47}$ \lum~and there is only one (J0428.6$-$3788 or PKS 0426$-$380, $z=1.11$) in the $\log L_{\rm BAT}=46.5-47$ bin and hence not shown. Number shown in brackets refer to number of sources in the respective luminosity bin. See the text for details.} \label{fig:lum_SED}
\end{figure*}

\section{Comparison with \fermi~Blazars}\label{sec6}
It may be of a great interest to compare the accretion and jet powers of BAT blazars with that detected by \fermi-LAT sources. The powers that the jet carries in the form of the bulk motion of electrons ($P_{\rm e}$), `cold' protons ($P_{\rm p}$, assuming one proton per electron, i.e. no pairs), radiation ($P_{\rm r}$), and magnetic field ($P_{\rm m}$) are calculated following \citet[][]{2008MNRAS.385..283C}. The derived jet powers, assuming a two-sided jet, are reported in Table~\ref{tab:jet}. For a comparison, we use the results reported for a large sample of \fermi~blazars in \citet[][]{2014Natur.515..376G} and \citet[][]{2017ApJ...851...33P}.

{\it $P_{\rm r}$ versus $L_{\rm d}$:} The most robust estimate of the jet power that we can have is the one that the jet produces in the form of radiation. This is because $P_{\rm r}$ is directly proportional to the observed bolometric luminosity. In the left panel of Figure~\ref{fig:bat-lat}, we show the behavior of $P_{\rm r}$ as a function of $L_{\rm d}$. The low $P_{\rm r}$-$L_{\rm d}$ end is mostly occupied by low-power HSP sources, whereas, luminous BAT FSRQs dominate at the high-end. We compute the partial Spearmann's coefficient $\rho$ \citep[][]{1992A&A...256..399P} and PNC to quantify the strength of the correlation and derive the parameters independent of the common redshift effect. The derived coefficients are $\rho=0.40\pm0.08$ with PNC $<$10$^{-10}$ and  $\rho=0.33\pm0.18$ with PNC $<$10$^{-10}$ for \fermi-LAT and \swift-BAT blazar populations, respectively. A positive $P_{\rm r}-L_{\rm d}$ correlation have earlier been reported for \fermi-LAT detected sources \citep[e.g.,][]{2014Natur.515..376G,2017ApJ...851...33P}, and our findings establishes the fact that \swift-BAT blazars also follow the same positive trend.

{\it $P_{\rm j}$ versus $L_{\rm d}$:} In the middle panel of Figure~\ref{fig:bat-lat}, we show the total jet power ($P_{\rm j}=P_{\rm p}+P_{\rm e}+P_{\rm m}$) versus $L_{\rm d}$. A positive correlation is observed which is statistically confirmed with $\rho=0.53\pm0.08$ with PNC $<$10$^{-10}$ and  $\rho=0.69\pm0.05$ with PNC $<$10$^{-10}$ for \fermi-LAT and \swift-BAT blazars, respectively. Similar to LAT blazars, BAT detected ones too host jets with $P_{\rm j}>L_{\rm d}$. Interestingly, at the high-end of the jet power ($P_{\rm j}>10^{48}$ \lum), BAT blazars seems to have more extreme jets compared to LAT blazars. This is probably a selection effect due to higher detection threshold of \swift-BAT which leads to  the identification of the most luminous sources. In a more physical scenario, the observational shift of the SED peaks to lower frequencies with increase in the bolometric luminosity \citep[e.g.,][]{1998MNRAS.299..433F} can also explain this observation. BAT blazars that likely have inverse Compton peak located at lower energies with respect to LAT detected ones, host more powerful jets and luminous accretion disks.

\section{Luminosity Dependent Evolution }\label{sec7}
It is instructive to determine the average SEDs of BAT blazars with the motivation to explore their luminosity dependent evolution \citep[][]{2012ApJ...751..108A}. We divide the sources in different BAT luminosity ($L_{\rm BAT}$) bins and also based on their SED appearances to identify FSRQs/BL Lacs. The latter is necessary since 105-month BAT catalog does not provide FSRQ/BL Lac classification. Moreover, we consider only those sources that have both \fermi-LAT and \swift-BAT detections. We take the arithmetic average of the broadband SEDs in each $L_{\rm BAT}$ bin and show in Figure~\ref{fig:lum_SED}. Note that the adopted choice of $L_{\rm BAT}$ bins are mainly driven to distinguish BAT blazars of various powers and does not carry any other physical meaning. 

In the left panel, we show averaged SEDs for FSRQs in four $L_{\rm BAT}$ bins and note the increasing dominance of inverse Compton peak (i.e. more Compton dominated SED) with increasing $L_{\rm BAT}$ and $L_{\rm d}$. Also, there is an indication for the shift of the inverse Compton peak to lower frequencies as their power increases, especially in the highest $L_{\rm BAT}$ bin. In the right hand panel of Figure~\ref{fig:lum_SED}, we plot the SEDs for BL Lacs. There are no BL Lacs in our sample with $L_{\rm BAT}>10^{47}$ \lum~and only one (J0428.6$-$3788 or PKS 0426$-$380, $z=1.11$) has $\log L_{\rm BAT}$ in the bin $46.5-47$ and hence not plotted. The lowest $L_{\rm BAT}$ bin is dominated by HSP BL Lacs and a pronounced shift of the SED peaks to lower frequencies can be observed in the higher luminosity bin. These results are aligned with our current understanding about the blazar sequence where the most powerful objects are LSP FSRQs with Compton dominated SEDs and the luminosity dependent SED evolution is more pronounced in BL Lacs \citep[see][for latest results]{2017MNRAS.469..255G}.

\section{Summary}\label{sec8}
In this work, we study 146 blazars detected in the first 105-month all-sky survey of \swift-BAT. We analyze the \fermi-LAT data and 14$-$195 keV BAT spectra and supplement them with multiwavelength archival spectral measurements. We determine the physical properties of the jet and central engine by modeling the broadband SED with a simple leptonic emission model. Our main findings are summarized below.

\begin{enumerate}
\item We $\sim$quadruple the number of \swift-BAT detected blazars and also number of \fermi-LAT undetected ones compared to earlier studies \citep[146 versus 38, 45 versus 12, respectively,][]{2009ApJ...699..603A}.
\item In the photon index versus 14$-$195 keV luminosity plane, we find tentative evidences ($\rho= -0.29\pm0.08$, PNC = 0.01) of inverse correlation with more luminous blazars tend to have flatter spectrum.
\item The overall physical properties of BAT blazars resemble well with that known for LAT blazar population studied in \citet[][]{2017ApJ...851...33P}.
\item The mass of the central black hole derived from the single-epoch optical spectroscopy matches reasonably well with that derived by modeling the optical-UV bump with a standard accretion disk.
\item BAT blazars exhibit a positive correlation between the power that the jet produces in the form of radiation and $L_{\rm d}$. Blazars with most luminous accretion disks host the most radiatively powerful jets.
\item Comparing the averaged SEDs of BAT blazars binned in different 14$-$195 keV luminosity bins, we find that more luminous sources have more Compton dominated SED. There is also a hint of the shift of SED peaks to lower frequencies as the power of the blazars increases and it is found to be more pronounced in low-luminosity BL Lac objects.
\end{enumerate}

\acknowledgments
We are grateful to the journal referee for a constructive criticism. M.K. acknowledges support from NASA through ADAP award NNH16CT03C. CR acknowledges support from the CONICYT+PAI Convocatoria Nacional subvencion a instalacion en la academia convocatoria a\~{n}o 2017 PAI77170080. K.O. is an International Research Fellow of the Japan Society for the Promotion of Science (JSPS, ID: P17321). We are grateful to the \fermi-LAT Collaboration internal referee Raniere Menezes for useful suggestions and Alberto Dom{\'{\i}}nguez, Judith Recusin and Philippe Bruel for critical reading of the manuscript. The \textit{Fermi} LAT Collaboration acknowledges generous ongoing support from a number of agencies and institutes that have supported both the development and the operation of the LAT as well as scientific data analysis. These include the National Aeronautics and Space Administration and the Department of Energy in the United States, the Commissariat \`a l'Energie Atomique and the Centre National de la Recherche Scientifique / Institut National de Physique Nucl\'eaire et de Physique des Particules in France, the Agenzia Spaziale Italiana and the Istituto Nazionale di Fisica Nucleare in Italy, the Ministry of Education, Culture, Sports, Science and Technology (MEXT), High Energy Accelerator Research Organization (KEK) and Japan Aerospace Exploration Agency (JAXA) in Japan, and the K.~A.~Wallenberg Foundation, the Swedish Research Council and the Swedish National Space Board in Sweden. Additional support for science analysis during the operations phase is gratefully acknowledged from the Istituto Nazionale di Astrofisica in Italy and the Centre National d'\'Etudes Spatiales in France. This work performed in part under DOE Contract DE-AC02-76SF00515.  We acknowledge the work of the \swift-BAT team to make this study possible.

This research has made use of data obtained through the High Energy Astrophysics Science Archive Research Center Online Service, provided by the NASA/Goddard Space Flight Center. This research has made use of the NASA/IPAC Extragalactic Database (NED), which is operated by the Jet Propulsion Laboratory, California Institute of Technology, under contract with the National Aeronautics and Space Administration. Part of this work is based on archival data, software or online services provided by the ASI Data Center (ASDC). 

\software{fermiPy \citep{2017arXiv170709551W}}.

\bibliographystyle{aasjournal}
\bibliography{Master}

\clearpage

\begin{table*}
\caption{\swift-BAT detected sources which were classified as `beamed AGN' in 105-month catalog but excluded in this work.\label{tab:excluded}}
\begin{center}
\begin{tabular}{cccc}
\hline
BAT Name & Counterpart name & Redshift & Comments \\
\hline
 J0048.8+3155       &Mrk 348&                   0.015   &   Seyfert 2 galaxy \citep{2015AnA...579A..90H} \\
 J0136.5+3906       &B3 0133+388                &--&        No redshift \\
 J0142.0+3922       &B2 0138+39B&               0.08    &   \fermi-LAT undetected HSP BL Lac object \\
 J0156.5$-$5303     &RBS 259                    &--&      	No redshift\\
 J0225.8+5946       &NVSS J022626+592753        &--&       	No redshift\\
 J0241.3$-$0816     &NGC 1052&                  0.005   & 	LINER galaxy \citep{2018MNRAS.480.1106C}\\
 J0245.2+1047       &4C +10.08&                 0.07    & 	\fermi-LAT undetected HSP BL Lac object\\
 J0319.7+4132       &NGC 1275&                  0.018   & 	Seyfert 1.5 galaxy \citep{2010AnA...518A..10V}\\
 J0433.0+0521       &3C 120&                    0.033   & 	Radio galaxy \citep{2017ApJ...838...16T}\\
 J0550.7$-$3215B    &2XMM J055054.3$-$321616    &--&       	No redshift, counterpart doubtful\\
 J0608.9$-$5507     &PKS 0607$-$549             &--&       	No redshift\\
 J0612.2$-$4645     &PMN J0612$-$4647           &--&       	No redshift\\
 J0632.1$-$5404     &1RXS J063200.7$-$540454&   0.204   & 	Giant radio quasar \citep{2005AJ....130..896S}\\
 J0640.3$-$1286     &PMN J0640$-$1253           &--&       	No redshift\\
 J0709.3$-$1527     &PKS 0706$-$15              &--&       	No redshift\\
 J0923.2+3850       &B2 0920+39                 &--&       	No redshift\\
 J1213.2$-$6020     &1RXS J121324.5$-$601458    &--&       	No redshift\\
 J1304.5$-$5651     &IGR J13045$-$5630&         0.051   &   Radio and \fermi-LAT undetected, Galactic plane \\
 J1312.1$-$5631     &2MASX J13103701$-$5626551  &--&      	No redshift\\
 J1325.4$-$4301     &Cen A&                     0.002   & 	Radio galaxy\\
 J1512.2$-$1053A    &NVSS J151148$-$105023      &--&      	No redshift\\
 J1557.8$-$7913     &PKS 1549$-$79&             0.150   & 	Radio galaxy \citep{2012ApJ...747...95G}\\
 J1655.0$-$4998     &CXOU J165551.9$-$495732&   0.058   & 	Radio and \fermi-LAT undetected, Galactic plane\\
 J1719.7+4900       &ARP 102B&                  0.024   & 	LINER/Radio galaxy \citep{2014AnA...572A..66P}\\
 J1734.9$-$2074     &NVSS J173459$-$204533      &--&       	No redshift\\
 J1742.1$-$6054     &PKS 1737$-$60&             0.41    & 	\fermi-LAT undetected HSP BL Lac object\\
 J1941.3$-$6216     &PKS 1936$-$623             &--&       	No redshift\\
 J2037.2+4151       &SSTSL2 J203705.58+415005.3 &--&       	No redshift\\
 J2056.8+4939       &RX J2056.6+4940            &--&       	No redshift\\
 J2117.5+5139       &2MASX J21174741+5138523    &--&       	No redshift\\
 J2209.4$-$4711     &NGC 7213&                  0.006   & 	Seyfert 1 galaxy \citep{1984ApJ...285..475H}\\
 J2303.1$-$1837     &PKS 2300$-$18&             0.128   & 	Radio galaxy \citep{1987MNRAS.227...97R}\\
\hline
\end{tabular}
\end{center}
\tablecomments{We exclude three HSP blazars, J0142.0+3922, J0245.2+1047, and J1742.1$-$6054, which are not detected with \fermi-LAT. This is because the lack of the \gm-ray spectrum leaves their inverse Compton emission completely unconstrained.}
\end{table*}

\begin{table*}
\caption{\swift-BAT detected sources which were not classified as blazars in 105-month catalog though exhibit broadband properties similar to beamed AGNs.\label{tab:added}}
\begin{center}
\begin{tabular}{cccc}
\hline
BAT Name & Counterpart name & Redshift & Old classification \\
\hline
J0144.8$-$2754 & PMN J0145$-$2733      &   1.155   &     Unknown AGN       \\
J0131.5$-$1007 & PMN J0131$-$1009      &   3.515   &     Unknown AGN       \\
J0201.0+0329   & [HB89] 0158+031       &   0.765   &     Unknown AGN       \\
J1810.0$-$6554 & PMN J1809$-$6556      &   0.18    &     Unknown AGN       \\
J1105.4+0200   & PMN J1105+0202        &   0.106   &     U3                \\
J1254.9+1165   & CGRaBS J1254+1141     &   0.87    &     U3                \\ 
J1153.9+5848   & RGB J1153+585         &   0.202   &     U3                \\ 
J0909.0+0358   & 1RXS J090915.6+035453 &   3.2     &     Sy1;broad-line AGN\\
J0250.8$-$3626 & 6dF J0250552$-$361636 &   1.536   &     Sy1;broad-line AGN\\
J1310.9$-$5553 & IGR J13109$-$5552     &   0.104   &     Sy1                \\
J1347.1+7325   & RGB J1346+733         &   0.29    &     Sy1                \\
J0042.9+3016B  & RX J0042.6+3017       &   0.141   &     Sy1                \\
J0506.6$-$1937 & 6dF J0506479$-$193651 &   0.094   &     Sy1                \\
J0547.1$-$6427 & CGRaBS J0546$-$6415   &   0.323   &     Sy1                \\
J1132.9+1019B  & [HB89] 1130+106       &   0.539   &     Sy1                \\
J1224.1+4073   & NVSS J122359+404359   &   0.096   &     Sy1                \\
J1238.4+5349   & RGB J1238+534         &   0.347   &     Sy1                \\
J1306.4$-$7603 & PMN J1307$-$7602      &   0.183   &     Sy1                \\
J2033.3$-$2253 & PKS 2030$-$23	       &   0.131   &     Sy1                \\
J2033.4+2147   & 4C+21.55	       &   0.173   &     Sy1                \\
\hline
\end{tabular}
\end{center}
\tablecomments{Last column reports the classification presented in \citet[][]{2018ApJS..235....4O}. Source type `U3' were those for which X-ray data was not existed at the time of publication and `Sy1' were those classified as Seyfert 1 galaxy.}
\end{table*}

\begin{table*}
\caption{The \gm-ray spectral parameters of \fermi-LAT detected BAT blazars.\label{tab:fermi}}
\begin{center}
\begin{tabular}{lcccccccccccccccc}
\hline
Name & $F_{\rm 0.1-300~GeV}$ & $\Gamma_{\rm 0.1-300~GeV}/\alpha$ & $\beta_{\rm 0.1-300~GeV}$ & TS \\
~[1] & [2] & [3]  & [4] & [5] \\ 
\hline
J0036.0+5951 & 38.20$\pm$1.92 &  1.72$\pm$0.02 & -- & 3405 \\
J0122.9+3420 & 3.38$\pm$0.80 &  1.54$\pm$0.11 & -- & 164 \\
J0144.8$-$2754 & 7.02$\pm$0.29 &  2.59$\pm$0.04 & -- & 912 \\
J0213.7+5147 & 3.59$\pm$0.51 &  1.91$\pm$0.09 & -- & 162 \\
J0218.0+7348 & 10.70$\pm$0.41 &  2.93$\pm$0.05 & -- & 841 \\
\hline
\end{tabular}
\end{center}
\tablecomments{The column information are as follows: Col.[1]: name of the BAT blazar; Col.[2]: 0.1$-$300 GeV energy flux in units of 10$^{-6}$ MeV cm$^{-2}$ s$^{-1}$; Col.[3]: 0.1$-$300 GeV photon index for the power law model or photon index at pivot energy for log parabola model; Col.[4]: the curvature parameter in log parabola model; and Col.[5]: test statistic derived from the likelihood fitting. \\
(This table is available in its entirety in a machine-readable form in the online journal. A portion is shown here for guidance regarding its form and content.)}
\end{table*}

\newpage
\begin{table*}
\caption{The SED parameters associated with the modeling of the broadband emission of BAT blazars.\label{tab:sed_param}}
\begin{center}
\begin{tabular}{lccccccccccccccccc}
\hline
Name & $z$ & $\theta_{\rm v}$ & $M_{\rm bh}$ & Type & $L_{\rm d}$ & $R_{\rm d}$ & $R_{\rm blr}$ & $\delta$ & $\Gamma$ & $B$ & $p$ & $q$ & $\gamma_{\rm min}$ & $\gamma_{\rm b}$ & $\gamma_{\rm max}$ & $U_{\rm e}$ & CD \\
~[1] & [2] & [3]  & [4] & [5] & [6]  & [7] & [8] & [9] & [10] & [11] & [12] & [13] & [14] & [15]& [16]&[17] &[18] \\ 
\hline
J0010.5+1057	&	0.09	&	6.0	&	8.70	& D & 45.15	&0.215	&0.038	&9.5	&	11	&	0.2	&2.2	&5.1	&4	&	286	&		3.0e+03		&-2.16	&6.2\\
J0017.1+8134	&	3.37	&	4.0	&	10.04	& D & 48.00	&0.421	&1.020	&12.2	&	8	&	2.2	&1.9	&4.5	&1	&	41	&		2.0e+03		&-1.68	&16.2\\
J0036.0+5951	&	0.08	&	3.0	&	0.00	& N &0.00	&0.115	&0.000	&13.6	&	8	&	0.1	&2.1	&3.7	&50	&	251243	&		1.5e+06		&-2.71	&0.4\\
J0042.9+3016B	&	0.14	&	3.0	&	8.90	& A & 45.00	&0.031	&0.032	&16.5	&	11	&	1.2	&1.5	&5.3	&1	&	20	&		1.3e+03		&-1.69	&247.2\\
J0057.0+6405	&	0.29	&	3.0	&	8.70	& A & 44.00	&0.120	&0.010	&18.6	&	15	&	1.1	&2.1	&4.6	&5	&	300	&		3.0e+03		&-2.25	&0.6\\

\hline
\end{tabular}
\end{center}
\tablecomments{The column contents are as follows: Col.[1] and [2]: name and redshift of the source; Col.[3]: viewing angle, in degrees; Col.[4]: log scale black hole mass, in units of M$_{\odot}$; Col.[5]: method used to derive $M_{\rm bh}$ and $L_{\rm d}$, A: assumed; D: disk fitting, O: optical spectroscopy. The used suffixes are as follows: B: BASS spectroscopy \citet[][Mejia-Restrepo et al. in prep.]{2017ApJ...850...74K}, G11:\citet[][]{2011MNRAS.414.2674G}, S12: \citet[][]{2012ApJ...748...49S}, S13: \citet[][]{2013ApJ...764..135S}), WU02: \citet[][]{2002ApJ...579..530W}, T12: \citet[][]{2012RMxAA..48....9T}, W04: \citet[][]{2004ApJ...615L...9W} , and N: not used; Col.[6]: log scale accretion disk luminosity, in \lum; Col.[7] and [8]: dissipation distance and the size of the BLR, respectively, in parsec; Col.[9] and [10]: the Doppler factor and the bulk Lorentz factor, respectively; Col.[11]: magnetic field, in Gauss; Col.[12] and [13]: spectral indices of the broken power-law electron distribution before and after the break energy ($\gamma_{\rm b}$), respectively; Col.[14], [15], and [16]: the minimum, break, and the maximum Lorentz factors of the emitting electron distribution; Col.[17]: the log scale particle energy density, in erg cm$^{-3}$; and Col.[18]: Compton dominance. Note that $M_{\rm bh}$ and $L_{\rm d}$ values are quoted as zeros for those blazars whose SEDs are modeled with synchrotron SSC processes only, i.e., without invoking EC mechanism.\\
(This table is available in its entirety in a machine-readable form in the online journal. A portion is shown here for guidance regarding its form and content.)}
\end{table*}

\begin{table*}
\caption{Various jet powers derived from the SED modeling.\label{tab:jet}}
\begin{center}
\begin{tabular}{lccccc}
\hline
Name &  $P_{\rm e}$ & $P_{\rm m}$ & $P_{\rm r}$ & $P_{\rm p}$ & $P_{\rm j}$\\
~[1] & [2] & [3]  & [4] & [5] & [6]  \\ 
\hline
J0010.5+1057	& 44.84	& 44.40	& 45.23	& 46.97	& 46.97\\
J0017.1+8134	& 45.62	& 46.59	& 47.35	& 48.13	& 48.14\\
J0036.0+5951	 & 43.46	& 42.68	& 44.26	& 44.23	& 44.31\\
J0042.9+3016B	& 43.61	& 44.06	& 44.25	& 46.20	& 46.21\\
J0057.0+6405	& 44.51	& 45.44	& 45.01	& 46.51	& 46.55\\
\hline
\end{tabular}
\end{center}
\tablecomments{The column contents are as follows: Col.[1] name the source; Col.[2], [3], [4], [5], and [6]: the electron, magnetic, radiative, proton, and total jet powers, respectively. Note that $P_{\rm j}$ = $P_{\rm e}$ + $P_{\rm m}$ + $P_{\rm p}$. All jet powers are evaluated for a two-sided jet.\\
(This table is available in its entirety in a machine-readable form in the online journal. A portion is shown here for guidance regarding its form and content.)}
\end{table*}

\begin{table*}
\caption{The list of 28 blazars which are modeled with synchrotron and SSC processes, i.e., without invoking EC mechanism.\label{tab:syn}}
\begin{center}
\begin{tabular}{lc}
\hline
Name & $z$ \\
\hline
J0036.0+5951	 & 0.08	\\
J0122.9+3420	& 0.27	\\
J0213.7+5147	& 0.05	\\
J0232.8+2020	& 0.14\\	
J0244.8$-$5829	& 0.26\\	
J0326.0$-$5633	& 0.06	\\
J0349.2$-$1159	 & 0.18	\\
J0353.4$-$6830	& 0.09\\	
J0507.7+6732	 & 0.31	\\
J0550.7$-$3212A	& 0.07\\	
J0710.3+5908	 & 0.12	\\
J0721.0+7133	& 0.30	\\
J0733.9+5156	 & 0.07	\\
J0930.1+4987	 & 0.19	\\
J0934.0$-$1721	& 0.25\\	
J1031.5+5051	& 0.36	\\
J1103.5$-$2329	& 0.19	\\
J1104.4+3812	& 0.03	\\
J1136.7+6738	& 0.13	\\
J1221.3+3012	& 0.18	\\
J1417.7+2539	& 0.24	\\
J1428.7+4234	& 0.13	\\
J1943.5+2120	& 0.21	\\
J1959.6+6507	& 0.05	\\
J2009.6$-$4851 & 	0.07\\	
J2246.7$-$5208	& 0.19	\\
J2251.8$-$3210	& 0.25	\\
J2359.0$-$3038	& 0.17	\\
\hline
\end{tabular}
\end{center}
\end{table*}

\end{document}